\documentstyle[12pt,epsfig]{article}
\makeatletter
\@addtoreset{equation}{section}
\makeatother

\def\beq{\begin{equation}}
\def\eeq{\end{equation}}
\def\bea{\begin{eqnarray}}
\def\eea{\end{eqnarray}}

\def\del{\partial}
\def\eq#1{(\ref{#1})}
\def\Tr{{\rm Tr}}
\def\a{\alpha}

\def\G{\Gamma}
\def\d{\delta}
\def\th{\theta}
\def\r{\rho}
\def\s{\sigma}
\def\t{\tau}
\def\w{\omega}
\def\Im{{\rm Im} \, }

\def\beq{\begin{equation}}
\def\eeq{\end{equation}}
\def\ber{\begin{eqnarray}}
\def\eer{\end{eqnarray}}
\def\bea{\begin{eqnarray}}
\def\eea{\end{eqnarray}}
\def\del{\partial}
\def\vbar{\bar{v}}
\def\delbar{\bar{\del}}

\def\uhat{\hat{u}_n}

\def\a{\alpha}
\def\b{\beta}
\def\D{\Delta}
\def\l{\lambda}
\def\s{\sigma}
\def\t{\tau}
\def\vt{\vartheta}
\def\r{\rho}
\def\t{\theta}
\def\w{\omega}
\def\Tr{\mbox{Tr}}
\def\p{\psi}
\def\D{\Delta}

\def\del{{\partial}}

\newfont{\Bbb}{msbm10 scaled 1200}     %instead of eusb10
\newcommand{\mathbb}[1]{\mbox{\Bbb #1}}

\def\lbldef#1#2{\expandafter\gdef\csname #1\endcsname {#2}}

\def\href#1#2{#2}

\newcommand{\beqar}{\begin{eqnarray}}

\newcommand{\eeqar}{\end{eqnarray}}
\newcommand{\th}{\theta}

\def\del{\partial}
\def\vbar{\bar{v}}
\def\delbar{\bar{\del}}

\def\uhat{\hat{u}_n}

\def\a{\alpha}
\def\b{\beta}
\def\D{\Delta}
\def\l{\lambda}
\def\s{\sigma}
\def\t{\tau}
\def\vt{\vartheta}
\def\r{\rho}
\def\th{\theta}
\def\w{\omega}

\begin{document}

\baselineskip=15.5pt
\pagestyle{plain}
\setcounter{page}{1}
%\renewcommand{\thefootnote}{\fnsymbol{footnote}}
%--------+---------+---------+---------+---------+---------+---------+
%Title page
\begin{titlepage}

\leftline{\tt hep-th/0106129}

\vskip -.8cm

\rightline{\small{\tt CALT-68-2332}}
\rightline{\small{\tt CITUSC/01-20}}
\rightline{\small{\tt NSF-ITP-01-57}}

\begin{center}

\vskip 2 cm

{\LARGE {Open Strings on $AdS_2$ Branes}}

\vskip 2cm
{\large Peter Lee, Hirosi Ooguri, Jongwon Park,
 and Jonathan Tannenhauser}

\vskip 1.2cm

California Institute of Technology 452-48,
Pasadena, CA 91125

\vskip .7cm
{\tt peter, ooguri, jongwon, jetannen@theory.caltech.edu}

\vskip 1.5cm

{\bf Abstract}
\end{center}

\noindent

We study the spectrum of open strings on $AdS_2$ branes in
$AdS_3$ in an NS-NS background, using the $SL(2,R)$ WZW model. When
the brane carries no fundamental string charge, the
open string spectrum is the holomorphic square root of the spectrum of
closed strings in $AdS_3$. It contains short and long
strings, and is invariant under spectral flow.  When the brane carries
fundamental string charge, the open string spectrum again contains
short and long strings in all winding sectors.  However,
branes with fundamental string charge break half the spectral flow
symmetry.  This has different implications for short and long strings.  
As the fundamental string charge 
increases, the brane approaches the boundary of $AdS_3$. In this limit, 
the induced electric field on the worldvolume reaches its critical
value, producing noncommutative open string theory on
$AdS_2$.

\end{titlepage}

\newpage

%--------+---------+---------+---------+---------+---------+---------+
%Body

\section{Introduction}

The $SL(2,R)$ WZW model is ubiquitous in string theory.  It arises in
contexts ranging from the Liouville model in two-dimensional gravity
\cite{polyakov, gervais1, gervais2, kpz} to three-dimensional Einstein gravity
\cite{witten3d} to two-dimensional black holes \cite{witten2d} to 
Neveu-Schwarz 5-branes \cite{chs} and their relation to singularities
in Calabi-Yau spaces \cite{vm, vg, vo}.  In addition, the $SL(2,R)$ WZW model
describes the worldsheet of a string propagating in $AdS_3$ with a
background NS-NS $B$-field.  The application to string theory in
$AdS_3$ is of particular interest, as it opens a window onto
the AdS/CFT correspondence beyond the gravity approximation.  For all
of these reasons, the $SL(2,R)$ WZW model has been intensively
studied for more than a decade.\footnote{A representative sample of 
references is given in \cite{part2}.}  

Last year, a proposal was put forward 
\cite{part1} and checked \cite{part2} for the structure of the Hilbert
space of the model.  The symmetries of the theory require that its
Hilbert space decompose as a sum of irreducible representations of the
current algebra $\widehat{SL}(2,R) \times \widehat{SL}(2,R)$.  But which
representations appear, and with what multiplicities?  According to 
the proposal of \cite{part1}, the
Hilbert space contains discrete representations and continuous
representations, as well as their images under spectral flow.  It was
argued in \cite{part1} that the discrete representations and their
spectral flow images correspond to short strings in $AdS_3$, and the
continuous representations and their images to long strings.  In
both cases, the integer $w$ indexing the spectral flow was interpreted
as the winding number of the strings about the center of $AdS_3$.  

The analysis of \cite{part1} determined the spectrum of {\em closed} strings in
$AdS_3$.  We address the corresponding problem for {\em open} strings.  More
specifically, our setting is critical open bosonic string theory in
$AdS_3 \times {\cal M}$, with an NS-NS background, and in the presence of
a D-brane whose worldvolume fills an $AdS_2$ subspace of $AdS_3$ and
wraps some subspace of the compact space ${\cal M}$.  This {\em $AdS_2$
brane} preserves one linear combination of the left- and right-moving current
algebras.  Consequently, the Hilbert space of open strings ending on
the $AdS_2$ brane decomposes as a sum of irreducible representations
of a single $\widehat{SL}(2,R)$.  Our main task will be to determine
which representations appear in the spectrum.   

Some intuition may be gained from the $SU(2)$ counterpart of our
problem \cite{ks, ko, as, ars1, fffs, stanciu2, BDS, p, ars2}.  In
the $SU(2)$ case, the D-brane worldvolumes analogous to our $AdS_2$
branes are $S^2$ subspaces
embedded in the $SU(2)$ group manifold $S^3$.  These {\em $S^2$
branes} are quantized: if the level of the WZW model is $k$, there are $k+1$
possible D-brane configurations, labeled by a quantum number $n$
taking the values $n = -k/2, -k/2 +1 , \dots, k/2$.  If $k$ is even,
the D-brane with quantum number $n=0$ wraps the equatorial
$S^2$ within $S^3$.  In general, the D-branes associated with increasing $|n|$
wrap smaller and smaller 2-spheres.  By the time $n$ reaches $\pm
k/2$, the $S^2$ worldvolumes have degenerated to the north or south pole of
$S^3$.
  
The Hilbert space ${\cal H}_n$ of open
strings ending on the $S^2$ brane with quantum number $n$ decomposes as
\beq
\label{su2case}
  {\cal H}_n = \bigoplus_{j=0}^{k/2 - |n|} D_j \,,
\eeq
where $D_j$ is the irreducible representation of
the current algebra $\widehat{SU}(2)$ whose
ground states make up the spin-$j$ representation of
$SU(2)$ \cite{ars1}. 
It is evident from \eq{su2case} that the Hilbert space ``loses''
representations as $|n|$ increases.  For example, the Hilbert space
of open strings ending on an equatorial $S^2$ brane 
is ${\cal H}_{n=0} =  \bigoplus_{j=0}^{k/2} D_j$,
which is the holomorphic square root of the Hilbert space of 
closed strings in $SU(2)$, projected onto integral $j$. On the other hand, when $n=\pm
k/2$ and the D-brane has shrunk to  
the north pole or the south pole, the Hilbert
space is reduced to the single representation
${\cal H}_{n=\pm k/2} = D_{j=0}$.   

We seek a similarly detailed picture of the Hilbert space of  
$AdS_2$ branes in $AdS_3$.  Our method resembles that of
\cite{part1}: we start by constructing {\em classical} open string
worldsheet solutions, based on which we then conjecture
the form of the {\em quantum} Hilbert space. To test the validity of
this approach, and to introduce the tools we will need for $SL(2,R)$ 
in what may be a more familiar context, we begin in section
\ref{branes} with a semiclassical treatment of $S^2$ branes in the
$SU(2)$ WZW model. The Hilbert space structure that emerges from our
semiclassical methods is identical to that of (\ref{su2case}), though
we cannot see the quantization of the parameter $n$.\footnote{A semiclassical
argument for the quantization of $n$ was given in \cite{BDS}.} 
Our approach explains naturally in terms of the geometry of $S^3$ 
why $\widehat{SU}(2)$ representations are skimmed off the Hilbert
space as $|n|$ increases.

We then return to the $SL(2,R)$ WZW model.  Following a review in section
\ref{closed} of the closed string Hilbert space, we take up the
subject of $AdS_2$ branes in sections \ref{flatbrane} and
\ref{curvedbrane}.  Like the $S^2$ branes in $SU(2)$, the $AdS_2$
branes in $SL(2,R)$ are quantized.  The quantum number they carry is
essentially fundamental string charge.  In a suitable coordinate
system, each $AdS_2$ brane is located at some fixed value $\psi_0$ of
one of the coordinates.  The quantization condition is
\beq
\label{quantcond}
\sinh \psi_0 = g_s Q \,,
\eeq
where $g_s$ is the string coupling constant and  $Q$ is the
fundamental string charge carried by the brane. The condition
\eq{quantcond} restricts $\psi_0$ to a discrete (but now, neither
finite nor bounded!) set of allowed values.  As with the $S^2$ branes,
though, our analysis is insensitive to this quantization.

The simplest case, $\psi_0 = 0$, is treated in section
\ref{flatbrane}.  This case is the $SL(2,R)$ analogue of the
equatorial $S^2$ branes in the $SU(2)$ WZW model.  
An $AdS_2$ brane with $\psi_0 = 0$ is a ``straight'' brane cutting
through the middle of $AdS_3$.  The Dirichlet boundary condition
defining such a straight brane preserves the full spectral flow
symmetry of the closed string theory.
Semiclassical analysis suggests that the open string Hilbert space is
the holomorphic square root of the Hilbert space of closed strings in
$AdS_3$.  A one-loop Euclidean partition function calculation,
described in Appendix \ref{partition}, confirms
this conjecture.

Section \ref{curvedbrane} is devoted to branes with $\psi_0 \ne 0$.
Exact quantitative results are unavailable here; nevertheless, we
are able to arrive at a qualitative picture of the Hilbert space.    

An $AdS_2$ brane with $\psi_0 \ne 0$ is analogous to an $S^2$ brane with
$n \ne 0$ in the $SU(2)$ WZW model.  Varying $\psi_0$ away from zero 
curves the brane towards the boundary of $AdS_3$.  Unlike the
situation for $S^2$ branes in $S^3$, there is no loss of
representations as $|\psi_0|$ increases. This difference
is traceable to a simple difference in the geometry of the two setups.

Introducing an $AdS_2$ brane with $\psi_0 \ne 0$ breaks half
the spectral flow symmetry: the curved-brane Dirichlet boundary
condition is preserved only if the integer $w$ parametrizing the
spectral flow is even.  Nevertheless, we can construct classical
short and long string solutions---and the Hilbert space contains 
discrete and continuous representations---of both odd and even $w$.  
There is an important difference, though, between the short and long
string solutions, having to do with the action on $AdS_3$ of PT, the spacetime
parity and time-reversal symmetry.  When acting on 
discrete representations, PT flips the parity of $w$.
Thus it is possible to reach a discrete representation of any given 
value of $w$ from a discrete representation of any other given $w$
by a sequence of symmetry transformations: even $w$
spectral flow and, if necessary, target space PT.  Consequently, the 
$\psi_0$ dependence of the density of states of the discrete
representations is the same for all
$w$.  By contrast, PT maintains the parity of $w$ when acting on
continuous representations.  Thus the $\psi_0$ dependence of the
density of states of the continuous representations is different for
odd and even $w$.  We highlight this difference by examining the
contributions of the odd and even $w$ sectors to the divergence
structure of the one-loop Euclidean partition function.

In the limit $\psi_0 \rightarrow \pm \infty$,
the $AdS_2$ brane approaches the boundary of $AdS_3$,
and the induced electric field on the brane worldvolume approaches its 
critical value.  We therefore conjecture that the
$\psi_0 \to \infty$ limit reproduces noncommutative open string (NCOS)
theory on $AdS_2$. At the end of section \ref{curvedbrane}, we show that,
in this limit, the WZW Lagrangian takes a form similar to the Lagrangian
of noncommutative open strings  \cite{ncos, sst}, appropriately modified 
to account for the $AdS_2$ background.  We also take some
preliminary steps towards a computation of the one-loop partition function.

Section 6 contains a summary and some conclusions.

Towards the completion of this work, we received the preprint
\cite{petrib}, which contains some overlap with certain of our results.  
In addition, branes in $AdS_3$ were recently studied from a different
point of view in \cite{gks}.

\section{$S^2$ Branes in the $SU(2)$ WZW Model}
\label{branes}

In this section, we study $S^2$ branes in the $SU(2)$ WZW model 
from a semiclassical
point of view.  Our reasons for doing so are twofold.  First, the $SU(2)$ WZW
model is in several important ways similar to---and different
from---the  $SL(2,R)$ WZW model which is our main focus, and it is useful
to develop the ideas we will need later on in a more familiar setting.  
Second, the $SU(2)$ WZW model provides a testing ground for the
semiclassical techniques that will eventually help us maneuver through the
intricacies of the $SL(2,R)$ WZW model.  By examining
{\em classical} open strings ending on $S^2$ branes in $S^3$, we will be
led to the picture of \cite{ars1} for the structure of the
{\em quantum} Hilbert space.  

Before turning to the analysis of $S^2$ branes in $S^3$, let us begin 
with some remarks on branes in WZW models in general \cite{stanciu1, stanciu2}.
In free bosonic open string theory with target space
coordinates $X^a$, the standard boundary conditions at the string
endpoints may be expressed as
\beq
\label{freeglue}
\del_+ X^a = \pm \del_- X^a \,,
\eeq
with the plus sign indicating a Neumann condition and the minus sign a
Dirichlet condition.  In the free theory, the $\del_+ X^a$ ($\del_-X^a$) are
(anti-)holomorphically conserved currents.  The simplest extension to
strings propagating on group manifolds replaces $\del_+
X^a$ ($\del_-X^a$) by the (anti)-holomorphically conserved currents
$J^a_R$ ($J^a_L$).  In addition, we could generalize \eq{freeglue} to the
condition
\beq
\del_+ X^a = R^a_b \del_-X^b \,,
\eeq
where $R^a_b$ is a constant matrix; which directions are Neumann and
which Dirichlet are then determined
by the eigenvalues of $R^a_b$.  The corresponding operation
in the WZW model involves ``twisting'' the
condition relating the left- and right-moving currents by an
automorphism $R$ of the Lie algebra \cite{stanciu1, stanciu2},
\beq
\label{openglue}
J^a_R + R^a_b J^b_L = 0 \, .
\eeq
The statement that $R$ is a Lie algebra automorphism means that, for
all group generators $T^a$ and $T^b$, 
$[R(T^a), R(T^b)] = R([T^a,T^b])$.  The automorphism $R$ is further
required to preserve conformal invariance at the worldsheet boundary.  In
\eq{freeglue} we had a choice of sign; choosing the plus sign in
\eq{openglue} ensures that the boundary conditions preserve the
affine symmetry algebra.\footnote{We will not consider the more
general brane configurations that can be obtained by relaxing this
requirement.}  

The condition \eq{openglue} is called a {\em gluing condition}, and
implies the boundary conditions
satisfied by the target space coordinates. 
The gluing condition \eq{openglue} defines branes, whose
worldvolumes within the group manifold $G$ extend in the directions for
which the boundary conditions derived from \eq{openglue} are
Neumann.  The brane worldvolume containing a fixed element $g \in G$
can be represented as the ``twined'' conjugacy class
\beq
\label{conjugacy}
{\cal W}^r_g = \{ r(h) g h^{-1} : h \in G \},
\eeq
where $r$ is the group automorphism induced near the identity from the
Lie algebra automorphism $R$.\footnote{For $X$ in the Lie
algebra and $t$ sufficiently small, $r(e^{tX}) = e^{tR(X)}$.}   \

Now let us make this concrete for the case of branes in $S^3$, the
group manifold of $SU(2)$.  We write the general $SU(2)$ group element as 
\beq 
\label{para1}
g = \exp \left(i
\vec{\psi} \cdot \vec{\s} \right) \,, 
\eeq 
where 
\beq 
\vec{\psi} =
\left( \psi - {\pi \over 2} \right) (\sin \w \cos \phi, \sin \w
\sin \phi, \cos \w) \,, 
\eeq 
the coordinates $(\psi, \w, \phi)$
lie in the ranges 
\beq 
-{\pi \over 2} \le \psi \le {\pi \over 2} \,,
\qquad 0 \le \w  \le \pi \,, \qquad 0 \le \phi \le 2 \pi \,, 
\eeq and
$\vec{\s} = (\s_1, \s_2, \s_3)$ is the vector of Pauli matrices.
Explicitly, in this parametrization, 
\beq 
g = \left(
\begin{array}{cc} \sin \psi - i \cos \w \cos \psi & -i e^{-i \phi}
\cos \psi \sin \w \\ 
-i e^{i \phi} \cos \psi \sin \w & \sin \psi +
i \cos \w \cos \psi \end{array} \right) \,. 
\eeq  
The $S^3$ metric in these coordinates is
\beq
ds^2 = d\psi^2 + \cos^2  \psi \, (d\w^2 + \sin^2 \w \, d\phi^2) \,,
\eeq
from which it is apparent that the surfaces of constant 
$\psi$ are 2-spheres embedded in $S^3$.

Classical string worldsheets in $S^3$ are given by solutions of the
$SU(2)$ WZW model.  We take the worldsheet coordinates to be $\t$ and
$\s$.  For closed strings, $\s$ is periodic with period $2 \pi$; for
open strings, $0 \le \s \le \pi$.  It is also convenient to define the
light-cone combinations
\beq
x^\pm = \t \pm \s \,.
\eeq

The WZW theory possesses three right- and left-moving
currents, which may be grouped into the matrices
\beq 
J_R = k \, \del_+ g g^{-1} \,, \qquad J_L = k \, g^{-1} \del_- g \,; 
\eeq 
here $k$ is the level of the WZW model.
The WZW equations of motion state that these currents are conserved,
{\em i.e.}, $\del_- J_R = \del_+ J_L = 0$.

Taking the Lie algebra automorphism in \eq{openglue} to be trivial 
leads to the gluing condition 
\beq 
\label{GC}
J_L = - J_R \,.
\eeq 
The worldvolumes of the associated branes are ordinary conjugacy
classes of $S^3$: they are the 2-spheres given by 
\beq 
\Tr \, g = 2 \sin \psi_0 \,, 
\eeq 
for some constant $\psi_0 \in\left[-\pi / 2, \pi / 2\right]$. The 
worldvolume of the {\em $S^2$ brane} with $\psi_0 = 0$ 
spans the equatorial 2-sphere of $S^3$; as $|\psi_0| \to \pi/2$, the
$S^2$ branes degenerate to single points at the north or south pole.  

It is useful to introduce a second parametrization of the
$SU(2)$ group element, 
\beq 
\label{para2}
g = \left( \begin{array}{cc} \sin r
\sin \th + i \cos r \sin t & - \cos r \cos t - i \cos \th \sin r
\\ \cos r \cos t - i \cos \th \sin r & \sin r \sin \th - i \cos r
\sin t \end{array} \right) \,, 
\eeq 
where the coordinates $(r, \th,t)$ satisfy 
\beq 
0 \le r \le {\pi \over 2} \,, \qquad 0 \le \th,t \le 2 \pi \,. 
\eeq 
The $S^3$ metric in these coordinates takes the form
\beq
ds^2 = \cos^2 r \, dt^2 + dr^2 + \sin^2 r \, d\th^2 \,.
\eeq

One simple class of open string configurations satisfying the WZW 
equations of motion is given by 
\beq 
\label{simple}
t = a \t \,,\qquad  \th = a \s + \th_0 \,,\qquad  r =
r_0 \,, 
\eeq 
where $0 \le a \le 1$ and $0 \le \th_0 \le \pi/2$.\footnote{This range for 
$\th_0$ assumes 
that $\psi_0>0$.  If $\psi_0 < 0$, the proper range for $\th_0$ is
$\pi\le \theta_0 \le 3\pi/2$.  The bound $0 \le a \le 1$ will be
explained in greater detail in the analogous $SL(2,R)$ context
in section \ref{classicalcurved}; essentially, any solution with arbitrary  
$a$ can be mapped by spectral flow to a solution satisfying
the bound.  One important difference between the
$SL(2,R)$ and $SU(2)$ WZW models, though, is that, in the quantum theory
of the $SL(2,R)$ WZW model, spectral flow generates new representations of
the current algebra, whereas in the $SU(2)$ WZW model, it does not.}  
The solutions satisfy the Dirichlet condition defining the $S^2$
brane, provided 
\beq
\sin \th_0 = {\sin \psi_0 \over \sin r_0} \qquad \hbox{and} \qquad
a = 1 - {2 \th_0 \over \pi} \,. 
\eeq 
It follows that 
\beq 
\label{bound}
a \le 1 - {2 |\psi_0| \over \pi} \,. 
\eeq 
That is, $a$ has an upper bound
that decreases with increasing $|\psi_0|$.  The geometry of the
$S^2$ branes and their attached open strings is shown in Figure
1(a).

\begin{figure}[htb]
\label{su2brane}
\begin{center}
\epsfxsize=6.0in\leavevmode\epsfbox{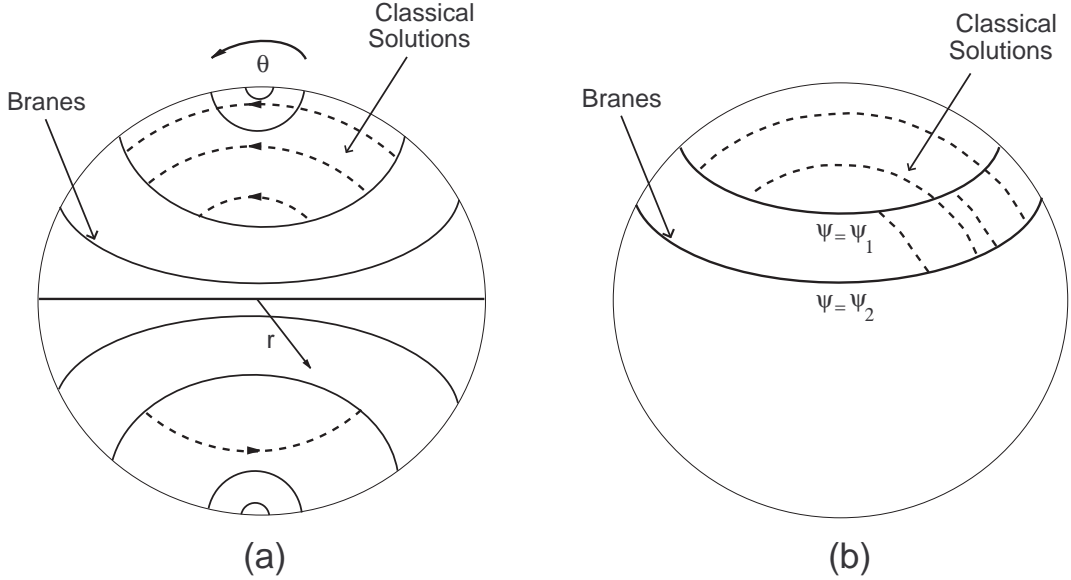}
\end{center}
\caption{(a) A view at fixed $t$ of $S^2$ branes and open strings
ending on them. (b) A view at fixed $t$ of a system of two $S^2$ branes
with open strings.}
\end{figure}

Let us compare the result of our classical analysis 
with the known structure of the quantum Hilbert space of open strings
ending on $S^2$
branes.  The Hilbert space is a sum of representations of
$\widehat{SU}(2)$;  the spin $j$ of a
representation appearing in the sum is related to the parameter $a$ of
the associated classical solution by $j = k a/2$.  
Although the analysis we have just presented is not refined enough to see it, 
$\psi_0$ is quantized \cite{BDS} as $\psi_0 = {\pi n \over k}$, where $n
= -k/2, -k/2 + 1, \dots, k/2$.  Thus, for given $\psi_0$, the bound
\eq{bound} on $a$ translates into a bound
\beq
\label{jbound}
j \le {k \over 2} - |n| 
\eeq
on the spins of the allowed representations in the Hilbert space.
This bound matches the conformal field theory analysis of \cite{ars1}, 
in which the Hilbert
space ${\cal H}_n$ of open strings ending on the brane labeled by $n$ 
was shown to be 
\beq
\label{decomp}
{\cal H}_n = \bigoplus_{j=0}^{k/2} N_{\a\a}^j D_j \,,
\eeq
where $\a = {1 \over 2}(n + {k \over 2})$, 
$D_j$ is the irreducible spin-$j$ highest weight representation of
$\widehat{SU}(2)$, and the fusion coefficients are given by
 \bea
\label{(con)fusion}
  N^j_{\a \b}=\left\{ \begin{array}{clc}
            1 & \qquad \mbox{if $|\a-\b|\le j \le$ min$\{\a+ \b,k-\a-\b\}$} \\
               & \qquad \mbox{and $2(j+\a+\b) \equiv 0$ mod $2$ }\,, \\
        0 & \qquad \mbox{otherwise}\,. \end{array}
        \right. 
\eea
{\em A priori}, the sum over $j$ is to be taken in half-integer steps
({\em i.e.}, $j= 0,1/2, \dots, k/2$).  However, the fusion coefficient
$N^j_{\a\a}$ is nonzero only if $j$ is an integer.  The equatorial
$S^2$ brane has $\a = k/4$; the branes at the poles have $\a = 0$ and $\a = k/2$.  For all $\a$, the sum cuts off at ${\rm
min} \, (2 \a, k - 2 \a)$, which is readily seen to be equivalent to
the cutoff \eq{jbound}.  Thus \eq{decomp} is identical to
\eq{su2case}. Our classical methods have reproduced 
information about the quantum Hilbert space. 

The restriction to integral $j$ in the sum \eq{decomp} (or
\eq{su2case}) can be simply understood, at least for equatorial $S^2$
branes. 
In addition to the stringy solutions \eq{simple}, the equatorial
$S^2$ brane also admits the particle-like geodesic solutions
\beq
\label{su2geo}
t = a \t \,, \qquad r = 0 \,. 
\eeq
Let us consider the $k \to \infty$ limit of the theory.  In this
limit, if we expand
around geodesic solutions like \eq{su2geo}, the WZW model reduces to
quantum mechanics on $S^2$.  Its Hilbert space is therefore the space
${\cal L}^2(S^2)$ of square-integrable functions on $S^2$.  This space
decomposes into spherical harmonics, which correspond to
representations of integral spin only.

We conclude this section with a slight generalization.  Let us
consider a system of two $S^2$ branes, located at $\psi = \psi_1 = \pi n_1 /k$
and $\psi = \psi_2 = \pi n_2 /k$; without loss of generality, we may
assume that $\psi_1 > 0$ and $|\psi_1| > |\psi_2|$, as shown in Figure
1(b). The
conformal field theory analysis that led to \eq{decomp} and the expression
\eq{(con)fusion} for the fusion coefficients now tells us that the Hilbert
space of strings stretching from the brane at $\psi_1$ to the brane at
$\psi_2$ has the decomposition
\beq
 {\cal H}_{n_1,n_2} =\bigoplus_{j={1 \over 2}|n_1-n_2|}^{{1 \over 2}
(k-n_1-n_2)} D_j \,.
\eeq
We wish to reproduce the bound on $j$ by semiclassical methods. We
consider classical solutions of the form \eq{simple}, subject to
the Dirichlet boundary conditions  
\bea
 \sin\theta_0\sin r_0 &=& \sin \psi_1 \,, \\
 \sin(a\pi+\theta_0)\sin r_0 &=& \sin \psi_2 \,.
\eea
An argument like the one in the single-brane example shows that now
$a$ is bounded both above and below,  
\beq
 {|\psi_1-\psi_2| \over \pi} \le a \le 1-{\psi_1+\psi_2 \over \pi} \,.
\eeq
The solutions saturating the inequalities are those with $r_0 = \pi/2$.  
Since $j=k a/2$, the bound on $a$ translates to
\beq
\label{j2}
  {1\over 2}|n_1-n_2| \le j \le {1\over 2}\left(k-n_1-n_2\right) \,,
\eeq
which matches the conformal field theory result.

\section{Closed Strings in $AdS_3$}
\label{closed}

Our search in sections \ref{flatbrane} and \ref{curvedbrane} for
solutions of the $SL(2,R)$ WZW model 
corresponding to open strings ending on branes will be
guided by the known properties of closed string solutions.  In
this section, we summarize the analysis of \cite{part1} of closed
bosonic string theory in $AdS_3$.  Our review has two parts.
First, we survey the classical closed string solutions of the
$SL(2,R)$ WZW model, beginning with simple geodesic solutions and
building up more complicated solutions by acting with global
isometries and spectral flow.  Next, we sketch the structure of
the quantum Hilbert space. 

\subsection{Classical Solutions}

The space $AdS_3$ is the group manifold of $SL(2,R)$.  If we think of
$AdS_3$ (with unit anti-de~Sitter radius) as the hyperboloid
\beq
\label{hyperboloid}
(X^0)^2 - (X^1)^2 - (X^2)^2 + (X^3)^2 = 1
\eeq
embedded in ${\bf R}^{2,2}$, then
a point in $AdS_3$ is given by the $SL(2,R)$ matrix
\beq
g = \left( \begin{array}{cc} X^0 + X^1 & X^2 + X^3 \\ X^2 - X^3 & X^0
- X^1 \end{array} \right) \, .
\eeq
As usual, to avoid closed timelike curves, we actually work with the 
universal cover of $AdS_3$.   We may alternatively parametrize $g$ in terms 
of the global coordinates
$(\r, \th, t)$ defined in Appendix A,
\beq
\label{element}
g = \left( \begin{array}{cc} \cos t \cosh \r - \cos \th \sinh \r &
\sin t \cosh \r + \sin \th \sinh \r \\ -\sin t \cosh \r + \sin \th
\sinh \r & \cos t \cosh \r + \cos \th \sinh \r \end{array} \right) \, .
\eeq
The $AdS_3$ metric in these coordinates is
\beq
ds^2 = - \cosh^2 \!\r \, dt^2 + d\r^2 + \sinh^2 \!\r \, d\th^2 \,.
\eeq
The parametrization \eq{element} is the $SL(2,R)$ counterpart of the
$SU(2)$ parametrization \eq{para2}.

Like the $SU(2)$ theory, the $SL(2,R)$ theory possesses three
conserved right- and left-moving currents,
\beq
J^a_R(x^+) = k \Tr \left(T^a \del_+ g g^{-1} \right) \,, \qquad
J^a_L(x^-) = k \Tr \left( T^{a*} g^{-1} \del_-g \right) \,  \quad \qquad (a =
+,-,3) \, .
\eeq
Here $k$ is again the level of the WZW model, and the $T^a$,
given in terms of the Pauli matrices by
\beq
T^3 = -{i \over 2} \s_2 \, ,\qquad T^\pm = {1 \over 2} (\s_3 \pm i
\s_1) \, ,
\eeq
form a basis of the Lie algebra of $SL(2,R)$. Sometimes we write the
currents in the matrix form
\beq
J_R = k \del_+ g g^{-1} \, , \qquad J_L = k g^{-1} \del_- g \,.
\eeq

The general solution $g$ of the WZW model can be factored as a product
of left-moving and right-moving parts,
\beq
\label{factored}
g(\s,\t) = g_+(x^+) g_-(x^-) \,,
\eeq
but, as we have said, it is useful to begin by studying geodesic
solutions, which depend only on the time coordinate $\t$.

The simplest timelike geodesic solution is
\beq
\label{pointpart}
g_0 = \left( \begin{array}{cc} \cos \, \a \t & \sin \, \a \t \\ -\sin
\, \a \t & \cos \, \a \t \end{array} \right) \,,
\eeq
describing a point particle at the center of $AdS_3$,
\beq
\label{pointpart2}
t = \a \t \, ,\qquad \r = 0 \,.
\eeq
The most general timelike geodesic can be obtained by acting on
\eq{pointpart} with the global isometry group $SL(2,R) \times SL(2,R)$
of the WZW model.  Such a solution has the form
\beq
\label{timelike}
g = U \left( \begin{array}{cc} \cos \, \a \t & \sin \, \a \t \\ -\sin
\, \a \t & \cos \, \a \t \end{array} \right) V \, ,
\eeq
where $U$ and $V$ are constant $SL(2,R)$ elements.  The parameter $\a$ in
\eq{pointpart} and \eq{timelike} is related through the Virasoro
constraints  to the conformal weight $h$ of the sigma model on the
compact space ${\cal M}$.
The conserved currents of the solution \eq{pointpart} are
\beq
J^3_L = J^3_R = {k \a \over 2} \, , \qquad J^\pm_L = J^\pm_R = 0 \, ,
\eeq
and the energy, defined as the sum of the zero modes of $J^3_L$ and
$J^3_R$, is $k \a$.

The construction of spacelike geodesics is similar.  The simplest
spacelike geodesic solution
\beq
\label{basicspacelike}
g_0 = \left( \begin{array}{cc} e^{\a\t} & 0 \\ 0 & e^{-\a\t} \end{array}
\right)
\eeq
describes a straight line through the spacelike section $t=0$ of
$AdS_3$. Its currents are
\beq
J^3_L = J^3_R = 0 \, , \qquad J^\pm_L = J^\pm_R = {k \a \over 2} \,,
\eeq
and its energy is zero. The most general spacelike geodesic solution is
\beq
\label{spacelike}
g = U g_0 V \,,
\eeq
where, again, $U$ and $V$ are constant $SL(2,R)$ isometries. 

Given a classical solution $\tilde g(\s, \t)  = \tilde g_+(x^+) \tilde
g_-(x^-)$,  we can
generate a new solution $g(\s,\t) = g_+(x^+) g_-(x^-)$ by {\em
spectral flow}, which involves setting
\beq
\label{spectralflow}
g_+ = e^{{i \over 2} w x^+ \s_2} \tilde g_+ \,, \qquad g_- = \tilde g_-
e^{{i \over 2} w x^- \s_2} \, ,
\eeq
for some integer $w$.  Spectral flow acts on the $AdS_3$ global
coordinates by
\beq
\label{flowedcoords}
\r \to \r \, , \qquad t \to t + w \t \, , \qquad \th \to \th + w \s \,
\, ,
\eeq
and on the $SL(2,R)$ currents by
\bea
J^3_R = \tilde J^3_R + {k w \over 2} \,, \quad && \quad  J^\pm_R = \tilde
J^\pm_R e^{\mp i w x^+} \,, \\ J^3_L = \tilde J^3_L + {k w \over 2} \,,
\quad && \quad J^\pm_L = \tilde J^\pm_L e^{\mp i w x^-} \, ;
\eea
or in terms of Fourier modes,
\beq
\label{flowedmodes}
J^3_{R,L \, n} = \tilde J^3_{R,L \, n} + {k w \over 2} \d_{n,0} \,, \qquad
J^\pm_{R,L \, n} =
\tilde J^\pm_{R,L \, n \mp w} \,.
\eeq

Timelike geodesics are mapped under spectral flow to short string
solutions, which expand and contract periodically in global time and
wind $w$ times around the center of $AdS_3$.  Spacelike geodesics are mapped to
long string solutions, which start in the infinite global-time past as
circular strings of infinite radius wound $w$ times near the boundary
of $AdS_3$, collapse (to a point, if there is no angular momentum) as
$t \to 0$, and expand again as $t \to \infty$ to wound circular
strings at the boundary.  After imposing the Virasoro constraints, the 
solutions constructed in this way have energies
\beq
\label{energy}
E = {k w \over 2} + {1 \over w} \left(\mp{k \a^2 \over 2} + 2 h \right)
\,,
\eeq
where the minus sign corresponds to short strings and the plus sign to
long strings.

\subsection{The Quantum Hilbert Space}

We now recall the structure of the quantum Hilbert space of the
$SL(2,R)$ WZW model.  The Hilbert space decomposes as the sum of {\em discrete}
representations of the $\widehat{SL}(2,R)$ current algebra, ${\em continuous}$
representations of the current algebra, and the images of these
representations under spectral flow.  Let us briefly review what this means.

The zero modes $J^a_0$ of the generators $J^a$ of the (left- or
right-moving) $\widehat{SL}(2,R)$ current algebra form a closed subalgebra,
which generates the group $SL(2,R)$.  Among the unitary
representations of this $SL(2,R)$ are the {\em principal
discrete} highest- and lowest-weight representations, which are
realized in the Hilbert space
\beq
{\cal D}_j^\pm = \{|j;m\rangle : m = \pm j, \pm (j+1), \pm (j+2),
\dots \} \,.
\eeq
The states $|j;m\rangle$ are simultaneous eigenstates of $L_0$, the
zeroth Virasoro generator obtained from the Sugawara construction, and
$J^3_0$, the $SL(2,R)$ Cartan generator, with eigenvalues $-j(j-1)$ and $m$.
These representations are unitary if $j > 0$.  The representations
$\hat {\cal D}_j^\pm$ of the $\widehat{SL}(2,R)$ current algebra are
constructed by considering the states in ${\cal D}_j^\pm$ as primary states for
the action of $J^a_n$.  That is, the states $|j;m\rangle$ are taken to
be annihilated by the $J^a_n$ with $n>0$, while the $J^a_n$ with $n<0$
are understood as creation operators, whose repeated action on the
states $|j;m\rangle$ yields states that fill out the representations 
$\hat {\cal D}_j^\pm$.  

Another unitary representation of $SL(2,R)$  
is the {\em continuous} representation, realized
in the Hilbert space
\beq
{\cal C}_j^\a = \{|j,\a;m\rangle: m = \a, \a\pm 1, \a\pm 2, \dots \}
\,,
\eeq
with $0 \le \a < 1$ and $j=1/2 + is$, for real $s$.  Again, the
states in $|j,\a;m\rangle$ are simultaneous eigenstates of $L_0$ and
$J^3_0$, with eigenvalues $-j(j-1)$ and $m$.  The representation
${\cal C}_j^\a$ of $SL(2,R)$ gives rise to a representation $\hat
{\cal C}_j^\a$ of the current algebra in the manner described above
for the discrete representations. 

As we have just noted, the representations $\hat {\cal D}^\pm_j$ and
$\hat {\cal C}^\a_j$ are described by the action of the current algebra modes
$J^a_n$ on their constituent states.  Spectral flow by $w$ units
alters the modes $J^a_n$, as noted in \eq{flowedmodes}, and so maps the
representations $\hat {\cal D}^\pm_j$ and $\hat {\cal C}^\a_j$ into new
representations  $\hat {\cal D}^{\pm,w}_j$ and $\hat {\cal C}^{\a,w}_j$.  The
closed string Hilbert space was proposed in \cite{part1} to be the direct
sum of $\hat {\cal D}^{+,w}_j \otimes \hat {\cal D}^{+,w}_j$ and
$\hat {\cal C}^{\a,w}_j \otimes \hat {\cal C}^{\a,w}_j$, summed over 
integers $w$.  The two factors in each tensor product account 
for left- and right-moving
states. The permissible values of $j$
for the discrete representations are bounded by ${1 \over 2} < j < {k-1
\over 2}$; note that, before the physical state conditions are
imposed, $j$ may be any real number in this range.
The states in  $\hat {\cal D}^{+,w}_j \otimes \hat 
{\cal D}^{+,w}_j$
correspond to wound short strings and their excitations,\footnote{The
representations $\hat {\cal D}^{+,w}_j$ and
$\hat {\cal D}^{-,w+1}_{{k \over 2} - j}$, may be identified.  This
accounts for the exclusion of $\hat {\cal D}^{-,w}_j$ from the list of
allowed representations: it would be redundant to include it.} while the
states in ${\cal C}^{\a,w}_j \otimes {\cal C}^{\a,w}_j$ correspond to
wound long strings and their excitations.  The spectra of both kinds
of strings were computed in \cite{part1} and were checked by an
independent calculation in \cite{part2}.

It is important to keep clear the distinction between the Hilbert
space of the WZW model and the Hilbert space of the physical string
theory.  The ($AdS_3$ part of the) latter is the subspace of the
former defined by the
Virasoro constraints.  Spectral flow is a symmetry of the WZW model,
but does not commute with the Virasoro constraints, and is therefore
not realized explicitly in the physical string Hilbert space.  While in our
analysis of classical solutions above and in subsequent
sections, we have freely borrowed and will continue to borrow the 
intuitions (and  language) of strings, we
should bear in mind that our conclusions really pertain to the Hilbert
space of the WZW model before the imposition of the Virasoro constraints. 

\section{The Straight $AdS_2$ Brane}
\label{flatbrane}

Having reviewed closed string theory in $AdS_3$, we now add
branes to the game.  We said in section \ref{branes} that the brane
worldvolumes in group manifolds can be thought of as twined conjugacy
classes of the form 
\beq
{\cal W}_g^r = \{r(h) g h^{-1} : h \in G \} \,,
\eeq
where $r$ is a group automorphism.   In the case $G=AdS_3$, it was
shown by Bachas
and Petropoulos \cite{bp} that the only physically allowed branes
are those for which $r$ is the nontrivial
outer automorphism that acts on a group element $h$, parametrized as a
$2 \times 2$ matrix as in \eq{element}, by
\beq
r(h) = \w_0^{-1} h \w_0 \, , \qquad \hbox{with} \qquad \w_0 = \left
( \begin{array}{cc} 0 & 1 \\ 1 & 0 \end{array} \right) \, .
\eeq
For this choice of $r$, the twined conjugacy class ${\cal W}^r_g$ is
equivalently characterized as the set of $g^\prime \in SL(2,R)$ such
that $\Tr \, (\w_0 g^\prime) = \Tr \, (\w_0 g)$.  The worldvolumes of
the resulting branes are two-dimensional, since they are parametrized
by arbitrary $SL(2,R)$ group elements subject to the single
condition
\beq
\label{dir}
\Tr \, (\w_0 g^\prime) = 2 \sinh \psi_0 \,,
\eeq
where $\sinh \psi_0$ is a constant whose physical meaning will become
apparent shortly.  The gluing conditions for the currents take the form
\beq
\label{gluing}
J_L = -\w_0 J_R \w_0 \,.
\eeq
Upon tracing through the procedure for
extracting boundary conditions for coordinates from the gluing
conditions, we indeed find the differential form of 
\eq{dir} as the Dirichlet condition for the coordinate transverse to the 
worldvolume, as well as the appropriate Neumann conditions for the 
coordinates parallel to the worldvolume.

Thinking of $AdS_3$ as the hyperboloid \eq{hyperboloid} in four-dimensional
space, \eq{dir} becomes the statement $X^2 = \sinh \psi_0$.  Subject to
this condition, \eq{hyperboloid} becomes
\beq
(X^0)^2 - (X^1)^2 + (X^3)^2 = 1 + \sinh^2 \psi_0 \, ,
\eeq
which is the equation of two-dimensional anti-de~Sitter space.  Thus the
worldvolume geometry of the physical branes of \cite{bp} is that of
$AdS_2$ embedded in $AdS_3$.  Analysis of the Dirac-Born-Infeld action
of these {\em $AdS_2$ branes} shows that they consist of one D-string
bound to some number of fundamental strings.  The fundamental string
charge $Q$ is proportional to the constant $\sinh \psi_0$; it follows that
$\psi_0$ is quantized. The quantization condition is
\beq
\sinh \psi_0 = g_s Q \,,
\eeq
where $g_s$ is the string coupling constant. 

In dealing with $AdS_2$ branes, it is convenient to switch to a
coordinate system in which the Dirichlet condition is simple.  The
{\em $AdS_2$ coordinates} $(\psi, \w, t)$ are defined in terms of the global
coordinates by
\beq
\label{ads2coords1}
\sinh \psi = \sin \th \sinh \r \, , \qquad \cosh \psi \sinh \w = -
\cos \th \sinh \r \, , \qquad t = t \, ,
\eeq
and in terms of the embedding hyperboloid coordinates by
\beq
\label{ads2coords2}
X^1 = \cosh \psi \sinh \w \, , \qquad X^2 = \sinh \psi \, , \qquad
X^0 + i X^3 = \cosh \psi \cosh \w e^{it} \,.
\eeq
The $AdS_3$ metric in $AdS_2$ coordinates is
\beq
\label{metric}
ds^2 = d\psi^2 + \cosh^2 \psi (-\cosh^2 \w dt^2 + d\w^2) \,.
\eeq

The $AdS_2$ coordinates are the $SL(2,R)$ analogue of the coordinate
system given in \eq{para1} for $SU(2)$. 
In $AdS_2$ coordinates, the Dirichlet condition giving the location of
the $AdS_2$ brane becomes $\psi = \psi_0$.  Some $AdS_2$ branes in $AdS_3$
are shown in Figure 2.

\begin{figure}[htb]
\label{ads2}
\begin{center}
\epsfxsize=3.5in\leavevmode\epsfbox{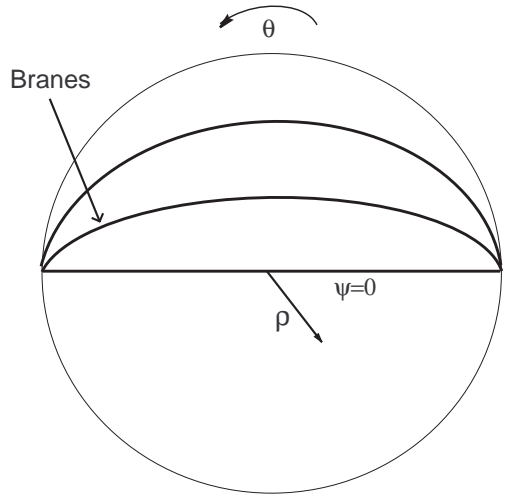}
\end{center}
\caption{$AdS_2$ branes in $AdS_3$.  The view is of the $(\r, \th)$ plane at 
fixed global time $t$.  The branes are surfaces of constant $\psi$.}
\end{figure}

In understanding $AdS_2$ branes and the strings that end on them,
there is a crucial distinction to be made between the ``straight'' branes
located at $\psi = \psi_0 =0$ and the ``curved'' branes located at $\psi =
\psi_0 \ne 0$. In this section, we study straight branes, following the
paradigm of section \ref{closed}.  First we construct classical geodesic
solutions confined to the brane. Next we investigate how spectral flow
generates classical string solutions that satisfy the appropriate
boundary conditions.  This leads to a proposal for the open string
spectrum.  We conclude by describing a check of this proposal by an explicit
partition function calculation modeled on that of \cite{part2}.

\subsection{Classical Solutions}
\label{classicalflat}

Our experience with closed strings suggests that looking at
geodesics might be a promising starting point for the study of open
string solutions.  A $\s$-independent solution that satisfies the
Dirichlet condition necessarily lies entirely within the brane.  The
timelike and spacelike geodesics \eq{pointpart} and \eq{basicspacelike}
obviously satisfy this requirement.  In the closed string case, we
built the most general timelike and spacelike geodesic solutions
from these basic ones by acting with the global isometry group
$SL(2,R) \times SL(2,R)$.  In the presence of an $AdS_2$ brane,
however, the only permitted isometries are those preserving the gluing
conditions
\beq
\label{glue}
J_L = - \w_0 J_R \w_0
\eeq
and leaving fixed the brane worldvolume---or equivalently, preserving the
Dirichlet condition
\beq
\label{dirichlet}
\Tr \, (\w_0 g) = 2 \sinh \psi_0 \,.
\eeq
The isometry $g \to U g V$ transforms the currents according to
\beq
J_R \to U J_R U^{-1} \, , \qquad J_L \to V^{-1} J_L V \,.
\eeq
The conditions for this isometry to preserve \eq{glue} and
\eq{dirichlet} are therefore
\beq
\label{iso1}
V^{-1} J_L V = -\w_0 U J_R U^{-1} \w_0 \,,
\eeq
\beq
\label{iso2}
\Tr \, (\w_0 U g V) = \Tr \, (\w_0 g) \,,
\eeq
which are satisfied if and only if $V = \w_0 U^{-1} \w_0$.  Thus the
boundary conditions imposed by the $AdS_2$ brane break the global
isometry group from $SL(2,R) \times SL(2,R)$ to a single $SL(2,R)$.
This is natural: the two factors of $SL(2,R)$ in closed string
theory correspond to independent transformations of left- and
right-moving modes, whereas the left- and right-moving modes of open
strings are related by the gluing conditions at the worldsheet
boundary. Our choice of gluing conditions guarantees that the
$SL(2,R)$ global symmetry is naturally promoted to an
affine symmetry, just as in the closed string case.

The preceding analysis of the breaking of the isometry group holds
for straight branes and curved branes alike: nothing in the argument
depended on the value of $\psi_0$.\footnote{If $\psi_0 = 0$, the most
general solution of \eq{iso1} and \eq{iso2} is $V = \pm \w_0
U^{-1} \w_0$, but the counting of parameters remains the same.}  
One difference between the
straight and curved cases is that, as we have noted, straight branes
contain particle-like solutions such as the ones shown in Figure
3.  Curved branes, on the other hand, do not. When we
generate the open string solutions associated with curved branes,
we will need a different starting point. We will explore this
in more detail in section \ref{curvedbrane}.

\begin{figure}[htb]
\label{fig12}
\begin{center}
\epsfxsize=5.0in\leavevmode\epsfbox{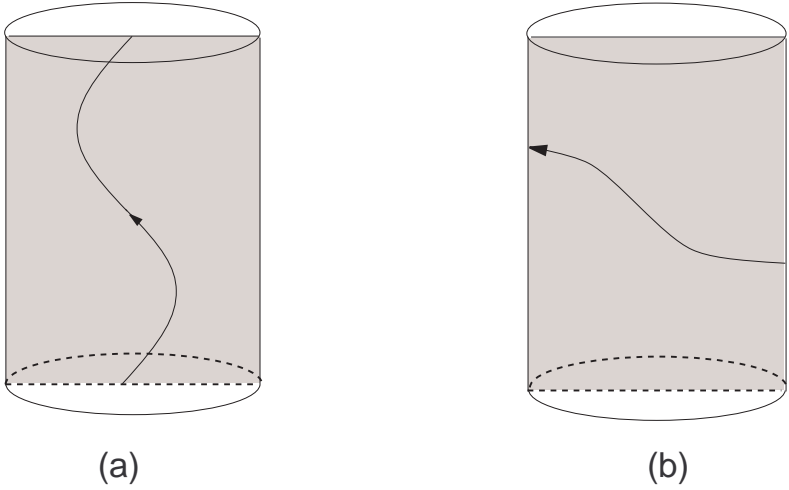}
\end{center}
\caption{(a) A timelike geodesic and (b) a spacelike geodesic confined
to the brane at $\psi_0=0$.}
\end{figure}

Spectral flow generates new open string solutions from old ones.  
In the closed string case, the parameter $w$ was required to be an
integer, to maintain the periodicity of the closed strings.  In the
presence of an $AdS_2$ brane, $w$ must again be integer, but for a
different reason: to ensure compatibility of spectral flow with the gluing
conditions \eq{glue}.

Is the Dirichlet condition compatible with spectral flow?  Suppose we
are given a solution satisfying the boundary condition
\beq
\sinh \psi \equiv \sin \th  \sinh \r  = \sinh \psi_0 \,
\eeq
at $\s = 0$ and $\s = \pi$.
If we act with $w$ units of spectral flow, we obtain a new solution
characterized by
\bea
\sinh \psi_{{\rm new}} = \sin \th \sinh \r =
\sinh \psi_0 \qquad && \hbox{at } \s = 0\,, \nonumber \\ \sinh \psi_{{\rm new}} = \sin
(\th + \pi w) \sin \r = \pm \sin \th \sinh
\r = \pm \sinh \psi_0 \qquad && \hbox{at } \s = \pi\,,
\eea
where the sign is plus if $w$ is even and minus if $w$ is odd. For a
straight $AdS_2$ brane, $\psi_0 = 0$, and so the Dirichlet condition
imposes no added restrictions on $w$.  On the other
hand, spectral flow is a symmetry of curved branes only for even
$w$.\footnote{If $w$ is odd, then a string whose $\s = 0$ endpoint
lies on a brane at  $\psi = \psi_0$ will end up at $\s = \pi$ with
$\psi = - \psi_0$. This will come in handy in section \ref{curvedbrane}.}
This is a key difference between straight and curved branes, and much of
section \ref{curvedbrane} will be devoted to its consequences.

\begin{figure}[htb]
\label{fig34}
\begin{center}
\epsfxsize=5.0in\leavevmode\epsfbox{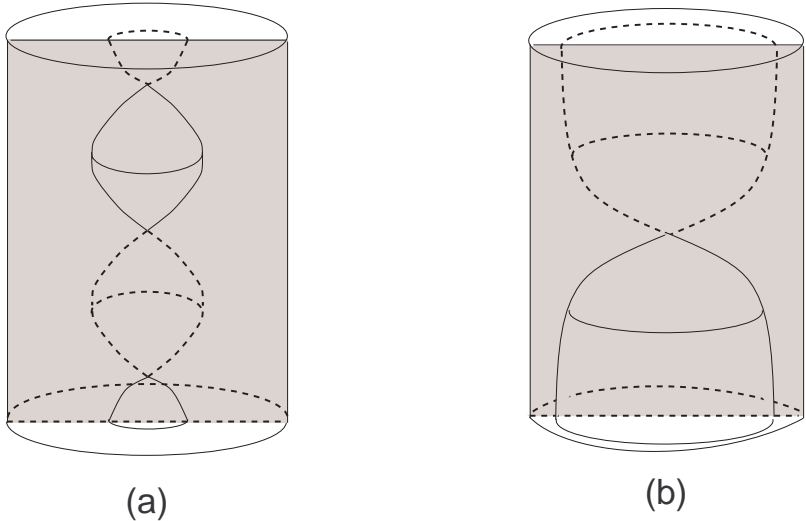}
\end{center}
\caption{(a) An open short string obtained from a timelike geodesic by 
spectral flow with $w=1$.  (b) An open long string obtained from a spacelike 
geodesic by spectral flow with $w=1$.}
\end{figure}

Spectral flow applied to timelike geodesics yields short strings. 
If $w$ is odd, these are
wound open strings that contract and expand periodically in $t$,
and whose endpoints are symmetric with  respect to the central
axis $\r=0$ of $AdS_3$. If $w$ is even, the string endpoints
coincide, giving wound circular strings. Spacelike
geodesics are mapped by spectral flow to long strings.  Examples
of strings of both kinds with $w=1$ are shown in Figure 4. A 
calculation just like the one sketched in section
\ref{closed} shows the spacetime energy of these solutions to be
\beq
\label{openenergy} 
E=J_0^3={kw\over 4}+{1\over w}\left(\mp{k\a^2\over
4}+h\right) \,, 
\eeq 
where the minus sign is for short strings and the plus sign for long
strings.  The energy of these open strings is precisely half the
energy \eq{energy} of their closed string counterparts.

\subsection{The Quantum Hilbert Space}
\label{quantumflat}

The classical solutions describing short and long strings ending
on the straight $AdS_2$ brane are in a sense ``exactly half'' of the
corresponding closed string solutions: left movers and right movers
are related by the gluing conditions, and the energy of the resulting
strings is half that of the closed strings.  As in the closed string
case, discrete representations ought to be associated with short
strings and continuous representations with long strings.  We
therefore propose that the quantum Hilbert space is the direct sum of
$\hat {\cal D}^{+,w}_j$ and $\hat {\cal C}^{\a,w}_j$, summed over all
integers $w$, and with ${1 \over 2} < j < {k -1 \over 2}$ for the
discrete representations and $j = {1 \over 2} + is$ with $s \in {\bf
R}$ for the continuous
representations.  Our proposed open string
spectrum is thus the holomorphic square root of the closed string
spectrum found in \cite{part1}. 

In Appendix \ref{partition}, we verify this conjecture by an
independent calculation of the spectrum.  We compute the
finite-temperature partition function in Euclidean $AdS_3$ and
interpret the result in terms of the free energy, summed over string
states.  This enables us to read off the spectrum.  
The result is in exact agreement with the conjecture.

\section{The Curved $AdS_2$ Brane }
\label{curvedbrane}

We now consider open strings constrained to end on a curved $AdS_2$
brane located at $\psi = \psi_0 \ne 0$; without loss of generality, we
may assume $\psi_0 > 0$. Our method is the same as in
the closed string and straight brane cases: we begin in section
\ref{classicalcurved} with the study of classical solutions
and their symmetries, and then in section
\ref{quantumcurved} conjecture the structure of the quantum
Hilbert space.  Several elements of the story are different for 
curved branes.  First, there are no classical geodesic
solutions.  Second, and more seriously, spectral flow is a symmetry of
the $AdS_2$ brane only if the winding number $w$ is even. Generating
solutions with odd winding number thus calls for new tricks, which we
describe in detail.

As we explain in section \ref{quantumcurved}, the chief consequence of these
differences is that, in the Hilbert space of the WZW model,
the density of states of the odd winding continuous representations
of $\widehat{SL}(2,R)$ behaves 
differently from the density of states of the even winding continuous 
representations as a function of the brane position
$\psi_0$.  Both kinds of representations are present in the spectrum for all 
$\psi_0$, as is the entire set of discrete representations.
This is to be contrasted with the $SU(2)$ WZW model, whose 
Hilbert space in the presence of $S^2$ branes loses representations as 
$\psi_0$ increases.  

In section \ref{betterthanone} we present a generalization to a system with two curved 
$AdS_2$ branes. In section \ref{NCOS} we study the limit $\psi_0 \to \infty$ in which the 
$AdS_2$ brane becomes highly curved.  In this limit, the WZW model Lagrangian 
resembles that of noncommutative open string theory in $AdS_2$.

\subsection{Classical Solutions}
\label{classicalcurved}

The program we followed in the last section began by constructing
$\s$-independent classical solutions lying within the brane.  If
$\psi_0 >0$, no such solutions exist.  For suppose $g = g(\t)$ is a
(non-constant) solution.  It must satisfy the Dirichlet condition $\Tr \,
(\w_0 g) = 2 \sinh \psi_0$, as well as the gluing condition \eq{glue},
which for the case at hand reads
\beq
\label{nogeodesic}
 g^{-1} \del_\t g = - \w_0 \del_\t g g^{-1} \w_0 \,,
\eeq
as $\del_+ g = \del_- g = {1 \over 2} \del_\t g(\t)$.  Multiplying on the
right by $(\del_\t g)^{-1}$, inverting, and taking the trace gives
$\Tr \, (g \w_0) = - \Tr \, (\w_0 g)$, and thus $\Tr \, (g \w_0) = 0$, in
contradiction with the Dirichlet condition.

Though there are no particle-like solutions, we are led to simple
string solutions by the following physical argument. Imagine
starting with the timelike geodesic \eq{pointpart2} in a flat
$AdS_2$ brane and then increasing $\psi_0$ by turning on a
background electric field on the brane.  The timelike geodesic on
the brane can be thought of as an infinitely small string,
whose endpoints are equally and oppositely charged with respect to
the background electric field.  As $\psi_0$ increases, the string
must stretch so that its tension will balance the forces due to the 
electric field.
This picture suggests the basic timelike string
solution shown in Figure 5(a), 
\beq 
\label{cylindricalarc}
t = \a \t \,, \qquad \th = \a \s + \th_0 \,, \qquad \r = \r_0 \,,
\eeq with \beq \label{even_cond} \th_0 = {\pi \over 2}
(1-\alpha)\,, \qquad
  \sinh\r_0 = \frac{\sinh\psi_0}{\sin\th_0} \,,
\eeq
and $0 \le \a < 1$.
\begin{figure}[htb]
\label{fig56}
\begin{center}
\epsfxsize=5.0in\leavevmode\epsfbox{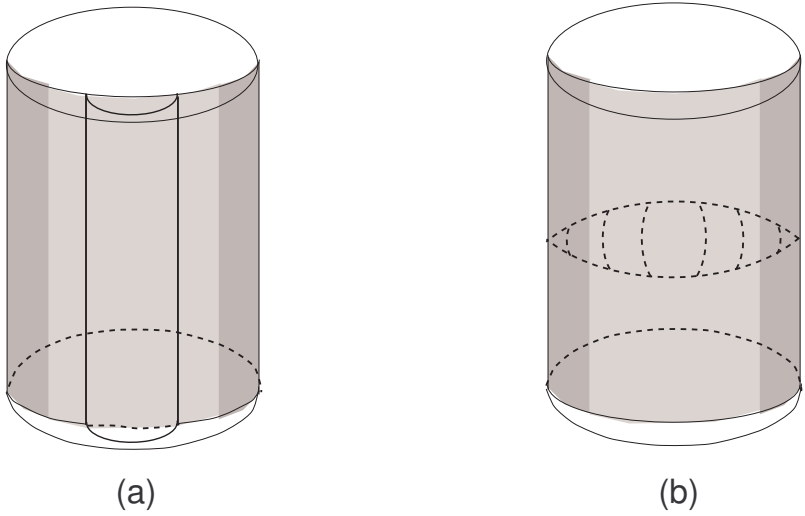}
\end{center}
\caption{The basic (a) ``timelike'' and (b) ``spacelike" string solutions ending on a
curved $AdS_2$ brane.}
\end{figure}
At fixed time, this solution describes a string curved in a circular arc 
and symmetric about $\th = \pi/2$.  It is easily checked that
\eq{cylindricalarc} solves the equations of motion.  Its
currents are the same as those of \eq{pointpart2}, and so it obeys
\eq{glue}. The choice of $\th_0$ and $\r_0$  ensure that
\eq{dirichlet} is satisfied as well.\footnote{The most trivial
solution imaginable is a ``pointlike instanton'': a solution with
$t$, $\r$, and $\th$ all constant.  It is amusing to observe that
\eq{cylindricalarc} may be obtained as the image under spectral flow
by $\a$ units of the pointlike instanton solution with $t = 0$, $\th
= \th_0$, and $\r = \r_0$.  Of course, $\a$ is, in general, fractional,
and so the notion of spectral flow here is purely formal.} 

The solution \eq{cylindricalarc} is the $SL(2,R)$ analogue of the $SU(2)$
solution \eq{simple} for open strings ending on $S^2$ branes.  An
important difference between the two is that the range of $\a$ in
\eq{cylindricalarc} is $0 \le \a < 1$, regardless of the value of
$\psi_0$, whereas the corresponding parameter $a$ in the $SU(2)$ case
is subject to the upper bound \eq{bound}, which depends on $\psi_0$.
This bound restricts the allowed representations in the $SU(2)$ WZW model
Hilbert space.  We will argue in section \ref{quantumcurved} that the
absence of such a bound for $AdS_2$ branes implies that there is no
similar restriction on the allowed representations in the $SL(2,R)$
model Hilbert space.    
Our classical analysis reveals this difference between the two theories to be
entirely geometric: in the $SL(2,R)$ WZW model, every $AdS_2$ brane
stretches from $\th = 0$ to $\th = \pi$, while in the $SU(2)$ case,
the range of $\th$, and hence of $a$, depends on $\psi_0$.  

The basic spacelike solution is given in matrix form as
\beq
\label{spacelikecurved}
g = \left( \begin{array}{cc} \sqrt{1+\b^2} \, e^{\a\t} & \b \, e^{\a(\s - {\pi \over 2})} \\
\b \, e^{-\a (\s-{\pi \over 2})}  & \sqrt{1+\b^2}\, e^{-\a\t}
\end{array} \right) \,,
\eeq
and in global coordinates as
\bea
\label{spacecoords}
\tan t &=& {\b \sinh(\a(\s - \pi/2)) \over \sqrt{1 + \b^2} \cosh \a \t}
\,, \nonumber \\
\tan \th &=& {- \b \cosh(\a(\s - \pi/2)) \over \sqrt{1 + \b^2} \sinh
\a \t} \,, \nonumber \\
\cosh^2 \r &=& (1 + \b^2) \cosh^2 \a \t + \b^2 \sinh^2(\a(\s -
\pi/2)) \,, 
\eea 
where $\b = \sinh\psi_0 / \cosh\frac{\pi}{2}\a$. This solution, depicted in
Figure 5(b), is a stringy generalization of the spacelike geodesic
\eq{basicspacelike}.\footnote{One way to
derive this solution is to assume first that $\psi_0$ is small and 
perturb the spacelike geodesic \eq{basicspacelike} accordingly.  From
the lowest-order corrections to \eq{basicspacelike} it is possible to guess
the form of \eq{spacelikecurved}, and to check that it is a valid
solution even if $\psi_0$ is not small.  The basic timelike solution 
\eq{cylindricalarc} may be derived from the timelike geodesic \eq{pointpart} 
by similar methods.}  It begins in the infinite worldsheet past
$\t = -\infty$ at $t = 0$ and at the edge $\th = 0$, $\r = \infty$ of
the brane, and arrives in the infinite worldsheet future $\t =
+\infty$ at the other edge $\th = \pi$, $\r = \infty$, again at
global time $t = 0$. Its excursion from the brane in the worldsheet interim
is governed by $\a$.  When $\a$ is small, the string stays near
the brane; as $\a$ increases, it strays farther and farther away.
A routine calculation shows that the currents of
\eq{spacelikecurved} are the same as those of \eq{basicspacelike}.

Having obtained the basic timelike and spacelike solutions, our next
task is to generate new solutions by acting with isometries and
spectral flow.  As we commented in section \ref{flatbrane}, the
allowed isometries are the same in the presence of curved and straight $AdS_2$
branes.  We have already seen, though, that
spectral flow is different.  If $w$ is even, spectral flow is still a
symmetry of the curved brane, and we can apply it to
\eq{cylindricalarc} and \eq{spacelikecurved} without incident.
For example, spectral flow applied to \eq{cylindricalarc} gives
\beq
t=(\alpha +w)\t \,, \qquad
\th
 = (\alpha +w)\s + \th_0 \,, \qquad
\r = \r_0  \,.
\eeq
If $w$ is even, this solution describes a cylindrical worldsheet making $w/2$
(not $w$!) complete cycles around the center of $AdS_3$, and whose endpoints
coincide with the endpoints of the original solution \eq{cylindricalarc}.

Applying an even amount of spectral flow to \eq{spacelikecurved}
gives a long string-like solution. Its properties are most easily seen
by replacing $t \to t + w \t $, $\th \to \th + w \s$ in  the
coordinate description \eq{spacecoords}. It begins in the infinite
spacetime past as a circular string of infinite radius with its
endpoints coincident at $\th = 0$.  Next it collapses to finite
radius. Its endpoints become separated in $t$ and move in towards
the center ($\th = \pi/2$) of the brane.  Finally, the string
re-expands towards infinite
radius, where its endpoints reconverge at $\th = \pi$.

How shall we generate solutions with odd $w$?  Our strategy will be
different in the timelike and spacelike cases.  The construction in
the timelike case makes use of the discrete target space symmetry
\beq
{\rm PT}: g \to \w_0 g \w_0 \,.
\eeq
Calling this symmetry PT is justified by its action on the global coordinates,
\beq
t \to -t \, , \qquad \th \to \pi - \th \, ,
\eeq
which reveals it as the composition of a parity and a time-reversal
transformation.  The PT symmetry acts on the currents by
\beq
J_{R,L} \to \w_0 J_{R,L} \w_0 \,,
\eeq
and hence on their modes by
\beq
J^3_{R,L \, n} \to -J^3_{R,L \,n} \,, \qquad J^\pm_{R,L \, n} \to -
J^\mp_{R,L \,n} \,.
\eeq
These expressions make it clear that PT is an automorphism of the
current algebra and a symmetry of the WZW model.  Moreover, it
preserves the gluing condition \eq{glue} and the Dirichlet condition
\eq{dirichlet}.

The PT symmetry maps short strings of winding number $w$ to short strings of
winding number $-w -1$.  To see this in a simple example, consider the
closed string or straight brane timelike geodesic \eq{pointpart2}. Acting
with $w$ units of spectral flow gives the solution
\beq
\label{flow}
t = (\a + w) \t \,;
\eeq
thus, without loss of generality, we may take $0 \le \a < 1$ in
\eq{pointpart2}, and regard a solution $t = \a \t$ with general
$\a$ as the image of the solution with $0 \le \a < 1$ under a suitable amount
of spectral flow.  Now if we apply PT to \eq{flow}, we get a
solution with
\beq
t = -(\a + w) \t = (\a' + (-w-1)) \t \,,
\eeq
where $\a' = 1 - \a$ satisfies $0 \le \a' < 1$.  By our previous
logic, this is to be thought of as the image of \eq{pointpart2} with
parameter $\a'$ under $-w-1$ units of spectral flow.

Our course for finding short string solutions with odd $w$ is now
clear: we simply act with PT on a solution with $w = 0$ ({\em
e.g.}, the image under an isometry of the basic timelike solution
\eq{cylindricalarc}) to reach the $w=-1$ sector, and then act on
the result with $w+1$ units of spectral flow, which is a symmetry
because $w+1$ is even.  As an example, if we implement this
procedure on \eq{cylindricalarc}, we find the solution \beq t = (1
- \a + w) \t \,, \qquad \th = (1 - \a + w) \s + \pi - \th_0 \,,
\qquad \r = \r_0 \,. \eeq Figure 6 shows a more 
complicated open short string solution with $w=1$.
\begin{figure}[htb]
\label{shortpsi}
\begin{center}
\epsfxsize=1.5in\leavevmode\epsfbox{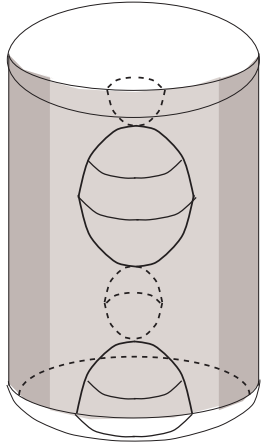}
\end{center}
\caption{A classical open $w=1$ short string solution.}
\end{figure}

This trick fails for long string solutions. An argument like
the one given above demonstrates that PT maps long string
solutions with winding number $w$ to long strings with winding
number $-w$, and therefore does not mix odd and even winding
sectors.  Instead, to construct spacelike solutions with odd $w$,
we recall that spectral flow leaves fixed the $\s=0$ endpoint
of  a string ending at $\psi = \psi_0$, but maps the $\s = \pi$
endpoint  to $\psi = -\psi_0$. This prompts us to introduce a
second $AdS_2$ brane located at $\psi = - \psi_0$.  If we can find
an unwound spacelike string with one endpoint on the brane at
$\psi_0$ and the other endpoint on the brane at $-\psi_0$, then
the action of spectral flow with odd $w$ will produce a string
with odd winding number, both of whose endpoints lie on the
$\psi_0$ brane.

Given \eq{spacelikecurved}, it is relatively straightforward to find unwound
spacelike strings stretching from the $\psi_0$ brane to the $-\psi_0$
brane.  There are two distinct classes of solutions, depending on the value of
$\a$.  Let $\b = \sinh \psi_0 / \sinh {\pi \over 2} \a$.  For  $|\b| < 1$, we have the solution
\beq
\label{b<1}
g = \left( \begin{array}{cc} \sqrt{1-\b^2}\, e^{\a\t} & -\b\, e^{\a(\s - {\pi \over 2})} \\
\b\, e^{-\a (\s-{\pi \over 2})}  & \sqrt{1-\b^2}\, e^{-\a\t} \end{array} \right) \, .
\eeq
For $|\b| > 1$,
\beq
\label{b>1}
g = \left( \begin{array}{cc} \sqrt{\b^2 -1}\, e^{\a\t} & -\b \, e^{\a(\s - {\pi \over 2})} \\
\b \, e^{-\a (\s-{\pi \over 2})}  & -\sqrt{\b^2-1}\, e^{-\a\t} \end{array} \right) \, .
\eeq

What do these solutions look like?  The two branes meet at the two
lines $\th = 0$ and $\th = \pi$ on the boundary $\r = \infty$
of $AdS_3$. The solution with $|\b| < 1$ describes a string that
begins in the infinite worldsheet past at the point $t=0$ on one
of these lines ($\th = \pi$ if $\b$ is positive), fills out a
spacelike surface between the two branes, and contracts in the
infinite worldsheet future to the point $t = 0$ on the other line.
The solution with $|\b| > 1$ describes a string that begins at a
point on one of the lines ($\th = \pi$ if $\b$ is positive), fills
out a timelike surface between the two branes with minimum radius 
$\cosh^{-1} \b$, and returns after an interval $\pi$ of
target space time to a point on the same line. The two cases are
sketched in Figure 7.
\begin{figure}[htb]
\label{fig78}
\begin{center}
\epsfxsize=5.0in\leavevmode\epsfbox{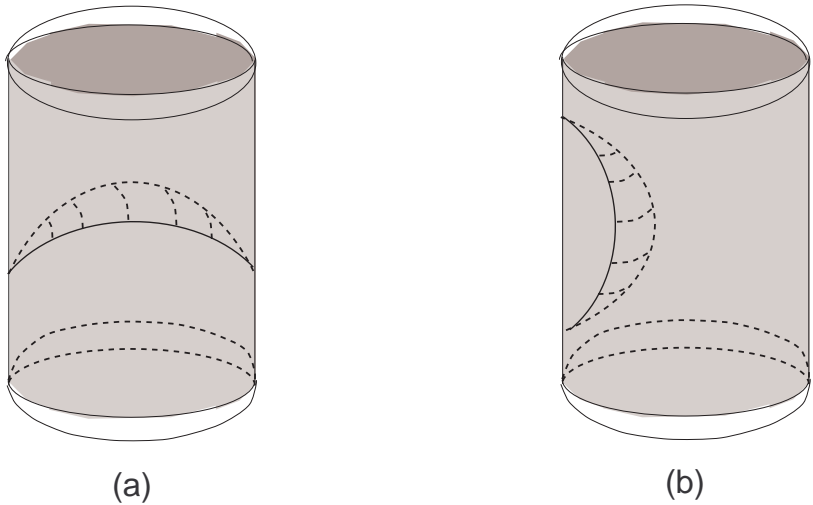}
\end{center}
\caption{Strings stretched between the branes at $\psi = +
\psi_0$ and $\psi = - \psi_0$, with (a) $|\b| < 1$ and (b) $|\b| > 1$.}
\end{figure}
The solution in the borderline case $|\b| = 1$, \beq
g =  \left( \begin{array}{cc} 0 & - e^{\a(\s - {\pi \over 2})} \\
 e^{-\a (\s-{\pi \over 2})}  & 0 \end{array} \right) \,,
\eeq
stretches between the centers $\th = \pi/2$ of the two branes at
global time $t = \pi/2$.  

Acting with an odd amount $w$ of spectral flow on these strings
gives long string solutions whose endpoints both lie on the
brane at $\psi = \psi_0$.  The image of \eq{b<1} under spectral
flow begins in the infinite spacetime past as a string of infinite
radius whose two endpoints lie on opposite edges of the brane.
With increasing $t$, the string contracts until, at $t=0$, the
endpoints cross at the center $\th = \pi/2$ of the brane.
Afterwards, the string expands until $t = \infty$, when the
endpoints again reach opposite edges of the brane.   At $t =
\infty$, each endpoint is at the edge opposite to the edge at
which it began at $t = -\infty$.  By contrast, in the flowed
solution with $|\b| > 1$, the endpoints do not have enough energy
to reach the center of the $AdS_2$ brane, and return at $t =
\infty$ to the edge at which they began.  Figure 8
depicts the long strings.

\begin{figure}[htb]
\label{fig910}
\begin{center}
\epsfxsize=5.0in\leavevmode\epsfbox{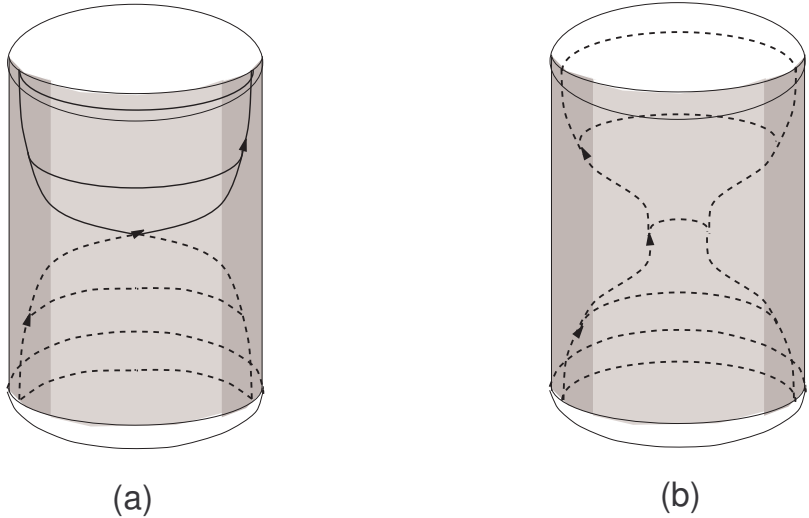}
\end{center}
\caption{A classical long string in the $w=1$ sector, for (a) $|\beta|<1$ 
and (b) $|\beta|>1$.}
\end{figure}

To summarize, we have constructed classical solutions for open strings
ending on a curved $AdS_2$ brane.   We began with simple timelike and
spacelike string solutions akin to the closed string and straight brane
geodesics considered in sections \ref{closed} and \ref{flatbrane}.  Acting
with the isometry group $SL(2,R)$ generated more unwound solutions.  As
in the straight brane case, spectral flow with even $w$ gave us new
winding solutions, but unlike the straight brane case, spectral flow with
odd $w$ was no longer a symmetry.  Short string and long string
solutions in odd winding sectors do exist, but to reach them, we
required new techniques: target-space PT symmetry for short
strings and the introduction of a brane at $\psi = -\psi_0$ for
long strings.  

\subsection{The Quantum Hilbert Space}
\label{quantumcurved}

In section \ref{quantumflat} we described the structure of the
Hilbert space of the $SL(2,R)$ WZW model in the presence of an $AdS_2$
brane at $\psi_0 = 0$.  Now, drawing on what we have learned about
classical open strings ending on curved branes, we sketch how the
Hilbert space changes as $\psi_0$ is increased above zero.

The WZW model Hilbert space at $\psi_0 = 0$ contains discrete 
representations $\hat {\cal D}_j^{+,w}$, for $w \in {\bf Z}$ and 
${1 \over 2} < j < {k-1 \over 2}$.  We propose that all of these
representations persist in the $\psi_0 > 0$ Hilbert
space. ``Discrete'' is something of a misnomer in the context of the 
WZW model of (the universal cover of)
$SL(2,R)$, since the discreteness of $j$ 
is enforced only by the Virasoro constraints.  At the
level of the WZW model Hilbert space, before the Virasoro constraints
are imposed, $j$ is not quantized, and it is meaningful to speak of
the density of states defined by a measure in $j$-space. 
In the presence of a  curved brane at $\psi = \psi_0$, this measure 
depends on $\psi_0$.  We conjecture that the $\psi_0$ dependence is
the same in all winding sectors.   

As in the straight brane case, the WZW model Hilbert space at nonzero 
$\psi_0$ contains continuous representations 
$\hat {\cal C}^{\a, w}_{{1 \over
2} + is}$, for all $\a \in [0,1)$, $w \in {\bf Z}$, and $s \in {\bf
R}$.  Once again, the density of states of these representations
now depends on $\psi_0$.  
We conjecture that the $\psi_0$ dependence
is the same in all even winding sectors and in all odd winding sectors,  
but that the dependence in the even winding sectors is different from the
dependence in the odd winding sectors.   

In support of our conjectures, we note first that, for all $\psi_0$,  
we were able to construct classical short string solutions for $\a$ 
satisfying $0 \le \a < 1$, which is the semiclassical version of the 
range ${1 \over 2} < j < {k-1 \over 2}$.  This, we observed, is unlike
the situation in 
the $SU(2)$ WZW model, where the range of $j$ for which classical
solutions exist is bounded in terms of $\psi_0$.  In the $SU(2)$ WZW
model, the truncation of
classical solutions as $\psi_0$ increased
manifested itself in the quantum theory as a loss of representations
in the Hilbert space.  In the $SL(2,R)$ WZW model, there is no
truncation classically, which leads us to believe that in the quantum
theory, representations spanning the entire range of $j$ are likewise
present.\footnote{A more computational argument in support of this
claim is presented in section \ref{NCOS}.} 
 The measure in the space of $j$ may depend on $\psi_0$, but
this is the only possible $\psi_0$ dependence in the structure of the
discrete sector of the Hilbert space.    

Spectral flow by an even amount is a symmetry of the theory.  It
follows that the $\psi_0$ dependence of the 
density of states must be the same in all
even winding sectors and in all odd winding sectors.  In addition, 
target space PT
symmetry links even and odd winding short string sectors.
Consequently, the  $\psi_0$ dependence of the density of discrete
states must in fact be the same in all winding sectors. 

The existence of classical long string solutions is strong evidence
that the WZW model Hilbert
space contains continuous representations.  Spectral flow by an even
amount is a symmetry of the long string
solutions, but target space PT symmetry does not mix even and odd
long string winding sectors.  Therefore, the $\psi_0$ dependence of
the density of states of the continuous representations is the same in
all even winding sectors and in all odd winding sectors, but the dependence in
the even sectors is, in general, different from the dependence in the
odd sectors.

We will presently provide further evidence that this is so by studying a
certain family of {\em physical} short string solutions in the even
and odd winding sectors.  We compute the energy $E$ of these solutions
as a function of their size $R$.  In both the odd and even winding
sectors, $E(R)$ increases monotonically at large $R$ to a value whose
functional form as a function of $w$ is the same in all sectors.
However, the dependence of $E$ on $R$---and on $\psi_0$---is
different in the two types of sectors.  The existence of an upper 
bound for $E(R)$ in the short string sectors implies that, above 
this bound, a continuous representation appears.  The $\psi_0$ dependence of 
$E(R)$ at large $R$ is different for odd and even $w$.  
This implies that the $\psi_0$ dependence of the density of states of 
the emergent continuous representations is likewise different for odd and 
even $w$.   

The short string solutions under
consideration belong to the physical Hilbert space.  To apply our
conclusions to the WZW model Hilbert space, we reason ${\em a\ fortiori}$:
since the ($AdS_3$ part of the) physical string Hilbert
space is the WZW model Hilbert space after the imposition of the
Virasoro constraints, if continuous representations exist in the
physical Hilbert space, surely they must exist in the WZW model
Hilbert space.

We now fill in the details of this argument.  To construct the even
winding solutions, we begin with the basic unwound timelike solution 
\beq
t = \a \t \,,\qquad \th = \a \s + \th_0 \,,\qquad \r = \r_0 \,,
\eeq
with
$0 \le \a < 1$, $\th_0 = {\pi \over 2} (1 - \a)$, and 
\beq
\label{evenw}
\sinh \r_0 = {\sinh \psi_0 \over \cos {\pi \a \over 2}} \,.
\eeq
We then act with the isometry
given by $U = e^{{1 \over 2} \r_1 \s_3}$ and $V = \w_0 U^{-1} \w_0 = U$,
where $\r_1$ is a constant. Finally, we perform an even amount $w$ of 
spectral flow.  The currents of the resulting solution are
\bea
\label{current1}
J^3_R &=& {k \over 2} (\a \cosh \r_1 + w) \,, \\
\label{current2}
J^\pm_R &=& \pm {i k \over 2} (\a \sinh \r_1 \, e^{\mp i w x^+}) \,,
\eea
and similarly for $J^a_L$.  

To construct the odd winding solutions, we once again 
introduce a second $AdS_2$ brane
at $\psi = - \psi_0$, and consider the unwound short string 
\beq
t = \a \t \,,\qquad \th = \a \s + \th_0 \,,\qquad \r  = \r_0 \,,
\eeq
with $0 \le \a < 1$, $\th_0 = \pi (1 - {\a \over 2})$, and
\beq
\label{oddw}
\sinh \r_0 = {\sinh \psi_0 \over \sin {\pi \a  \over 2}} \,.
\eeq
This is a string stretching between the two branes.  Again, we act
with the isometry given by $U = e^{{1 \over 2} \r_1 \s_3}$ and $V = \w_0
U^{-1} \w_0 = U$, and perform $w$ units of
spectral flow on the result.  Since $w$ now is odd, both endpoints of
the resulting solution lie on the brane at $\psi = \psi_0$.  The
currents of this solution, too, are given by \eq{current1} and
\eq{current2}.

To obtain physical string solutions in $AdS_3 \times {\cal M}$, we
must impose the Virasoro constraints $T^{AdS}_{\pm \pm} + h = 0$, 
where $T^{AdS}$ is the energy-momentum tensor for the $AdS_3$ modes of
the string, and $h$ is the energy-momentum tensor for ${\cal M}$,
which we regard as a conformal weight for the sigma model on ${\cal
M}$.  The energy-momentum tensor is calculable in terms of the
currents \eq{current1} and \eq{current2}, and the resulting Virasoro
constraint expresses $\a$ in terms of $h$ \cite{part1},
\beq
\a = \a_\pm = -w \cosh \r_1 \pm \sqrt{w^2 \sinh^2 \r_1 + {4 h \over
k}} \,.
\eeq    
Choosing the branch $\a = \a_+$, the spacetime energy takes the form
\beq
\label{e}
E = J^3_0 = {k \over 2} \left( \cosh \r_1 \sqrt{{4 h \over k} + w^2
\sinh^2 \r_1} - w \sinh^2 \r_1 \right) \,.
\eeq
This expression is valid for both the even and odd winding
solutions.

The last ingredient we need for our argument is a precise notion of
the size of the string.  We define $R$ to be the maximum value of the
coordinate $\r(\s, \t)$.  By writing even and odd winding solutions in
the matrix form \eq{element}, it is straightforward to calculate that,
for both types of solutions, 
\beq
\cosh R = \cosh \r_0 \cosh \r_1 \,.
\eeq
In \eq{e}, we expressed the string energy as a function of $\r_1$, the
isometry parameter.  We now work towards rewriting $E$ as a function
of $R$, in the large $R$ limit.  

As $\r_1 \to \infty$, $\a = \a_+$ approaches $0$ according to
\beq
\a =  \frac{{4 h \over k} - w^2}{w \cosh \r_1 + \sqrt{w^2 \sinh^2 \r_1
+ {4 h \over k}}} \sim {1 \over  w} \left( {4 h \over k} - w^2 \right)
e^{-\r_1} \,.
\eeq
It follows from \eq{oddw} that for odd $w$, in this limit,  
\beq
\sinh \r_0 = {\sinh \psi_0 \over \sin {\pi \a \over 2}} \sim {\sinh
\psi_0 \over {\pi \a \over 2}} \sim {2 w \sinh \psi_0 \over \pi ({4 h
\over k} - w^2)} e^{\r_1} \,,
\eeq
and therefore
\beq
\cosh R = \cosh \r_0 \cosh \r_1 \sim {2 w \sinh \psi_0 \over \pi ({4 h
\over k} - w^2)} e^{\r_1} \cosh \r_1 \,,
\eeq
so that
\beq
\label{oddR}
R \sim 2 \r_1 + \log \left( {2 w \sinh \psi_0 \over \pi ({4 h
\over k} - w^2)} \right) \,.
\eeq
For even $w$, as $\r_1 \to \infty$, $\r_0 \to \psi_0$, and so
\beq
\cosh R \sim \cosh \psi_0 \cosh \r_1 \,,
\eeq
or
\beq
\label{evenR}
R \sim \r_1 + \log \cosh \psi_0 \,.
\eeq
Substituting \eq{oddR} and \eq{evenR} into \eq{e}
allows us to determine $E$ as a function of $R$.  We find that, for
odd $w$,
\beq
E(R) = {h \over w}  + {kw \over 4} - {4h  - k w^2 \over 2 \pi k w^2} e^{-R}
\sinh \psi_0  \,,
\eeq
while for even $w$,
\beq
E(R) = {h \over w}  + {k w \over 4} - {(4h - k w^2)^2 \over 4 k
w^3} e^{-2R} \cosh^2 \psi_0 \,, 
\eeq 
plus terms respectively of order $e^{-2R}$ and $e^{-4R}$ for odd and even $w$.

What we learn from this analysis is that, in both the odd and even
winding sectors, $E(R)$ has an asymptote at ${h \over w}  + {k w \over
4}$.  This asymptote signals the existence of continuous
representations.  Solutions
with energies below the asymptote are
bound states---short strings---trapped within a finite radius in
$AdS_3$. By contrast, solutions with energies above the asymptote are free to
escape to the boundary of $AdS_3$.  These are the long strings, which
inhabit continuous representations.  This interpretation is consistent
with the fact that the energy of physical long strings with winding number $w$
is bounded below, in the semiclassical limit of large $h$ and large $k$, 
by ${h \over w}  + {k w \over 4}$, the asymptotic value of $E(R)$ .  
We have thus shown that continuous
representations of every $w$ exist in the physical string Hilbert space;
hence {\em a fortiori} such representations exist in the WZW model
Hilbert space.

We can take this argument one step further to confirm that the $\psi_0
> 0$ WZW model Hilbert space not only contains continuous
representations of every $w$,
but within each winding sector contains representations $\hat {\cal
C}^{\a, w}_{j = {1 \over 2} + is}$ of every $s$.
The Virasoro constraint determines $\a$ in terms of $s$ and
$h$.  Thus, if $s$ were somehow quantized in the WZW model Hilbert
space, the physical long string spectrum at fixed $w$ and $h$ would be
quantized as well.  As we have seen, though, the physical long string
spectrum at fixed $w$ and $h$ is continuous.  Thus $s$ must not be
quantized in the WZW model Hilbert space; representations with
arbitrary $s \in {\bf R}$ must actually appear.

The approach of $E(R)$ to its asymptote is different as a function of
$R$ and $\psi_0$ in the even and odd winding sectors. This is consistent with
our claim that the $\psi_0$ dependence of the density of states in the
WZW model Hilbert space is different in the odd and even winding long
string sectors.
An additional piece of evidence for this point comes from an analysis in
the spirit of \cite{part2} and Appendix \ref{partition} of the
divergence structure of the
one-loop Euclidean partition function.  The divergences in question 
signal the presence of continuous representations; they
originate in the infinite volume factors that appear when long strings 
are subject to a flat potential.  Accordingly, it is sufficient to
consider the contribution to the functional integral of the large $\r$
region.  In global coordinates, the WZW action at large $\r$ takes the
form
\beq
S \sim \int d^2z \left( \del \r \bar \del \r + {1 \over 4} e^{2 \r}
|\bar \del (\th - it)|^2 + \cdots \right) \,.
\eeq

In the one-loop calculation, the worldsheet is taken to be cylindrical
and of Euclidean signature; the target space time is likewise
Euclidean. The worldsheet coordinate $z = \s + i \t$ is subject to the
periodicity
\beq
\label{ws.}
\t \sim \t + 2 \pi t_W \,,
\eeq
with $t_W$ the worldsheet modulus.
At finite temperature, the target space coordinates $t$ and $\th$
describe a torus: $\th$ is periodic by nature, and $t$ is periodically
identified with period equal to the inverse temperature $\b$; that is,
\beq
\label{ts.}
\th - i t \sim \th - i t + 2 \pi n + i \b m  \,,
\eeq
where $m$ and $n$ are integers.  

If $\r$ is assumed to be fixed at $\r_0$, 
the equations of motion constrain $\th - i t$ 
to be a harmonic map from the worldsheet to the target space.  The
general harmonic map from the
cylinder to the torus is of the form
\beq
\label{harmonic}
\th - i t = w \s + i b \t + \th_0 \,,
\eeq
where $w$, $b$, and $\th_0$ are real constants.  Matching the
periodicities \eq{ws.} and \eq{ts.} of the worldsheet and target space
sets $b = {\b m \over 2 \pi t_W}$.  The partition function thus receives
contributions from harmonic maps of the form \eq{harmonic}, for all
integers $m$.  As in Appendix \ref{spectrum}, it is sufficient for our
purposes to concentrate on the $m=1$ sector.

If it is to describe a legitimate open string configuration, the map 
\eq{harmonic} must
satisfy the gluing conditions and the Dirichlet condition.  A straightforward calculation shows that, in the presence of an
$AdS_2$ brane at $\psi = \psi_0$, the gluing conditions are satisfied
if one of two conditions holds:
\begin{enumerate}
\item $w = b = \displaystyle{\b \over 2 \pi t_W}$; or
\item $\th_0 = 0$ or $\pi$, and $w$ is an integer.
\end{enumerate} 

Suppose condition 2 is fulfilled.  Then the Dirichlet
condition $\sin \th_0 \sinh \r_0 = \sinh \psi_0$ can be satisfied only
if $\psi_0 = 0$.  In this case, there is a family of solutions of the form
\eq{harmonic} with the desired properties, indexed by the continuous
parameter $\r_0$.  The solutions \eq{harmonic} are
holomorphic---that is, functions of $z = \s + i \t$---if the
worldsheet modulus takes the special value
\beq
\label{special}
t_W = {\b \over 2 \pi w} \,.
\eeq
As explained in \cite{part2}, the functional integral suffers a
logarithmic divergence at this special value of $t_W$.  The
divergence is the hallmark of a continuous representation, whose
density of states can be derived by properly regularizing the
infinity.  We have thus arrived again at a conclusion we reached in
section \ref{flatbrane}: the straight brane Hilbert space contains
continuous representations for all values of the winding number.

If $\psi_0 \ne 0$, then harmonic maps satisfying the gluing conditions
must fulfill condition 1.  In this case, the Dirichlet condition reads
\beq
\label{dircond}
\sin \th_0 \sinh \r_0 = \sin (w \pi + \th_0) \sinh \r_0 = \sinh \psi_0
\,.
\eeq
If $w$ is not an integer, \eq{dircond} determines $\th_0$ and $\r_0$
uniquely.  If $w$ is an odd integer, \eq{dircond} has no solution.  If
$w$ is an even integer, \eq{dircond} collapses to the single condition 
$\sin \th_0 \sinh \r_0 = \sinh \psi_0$, which has a family of
solutions indexed by a single continuous free parameter.  Condition 1
trivially implies \eq{special}; thus the maps \eq{harmonic} are
holomorphic.  The functional integral thus has a divergence
at $t_W = {\b \over 2 \pi w}$ if $w$ is even, but this divergence is
apparently absent if $w$ is odd.  Following the logic of the last
paragraph, we conclude that the WZW model Hilbert space contains
continuous representations in the even winding sectors.  On the other
hand, this line of reasoning tells us nothing about the odd winding
continuous representations.  Of course, we have already independently 
established that continuous representations exist in all winding
sectors.  This argument points to a difference in the
mechanism for generating continuous
representations in the even and odd winding sectors,
illustrating our claim that the nature of the continuous representations at
nonzero $\psi_0$---and in particular, their density of
states---depends significantly on the parity of the winding number.  

\subsection {A Two-Brane System}
\label{betterthanone}

An interesting perspective on the Hilbert space in the presence of a
curved brane at $\psi = \psi_0$ is obtained by revisiting a device from section
\ref{classicalcurved}: the introduction of a second brane at $\psi = -
\psi_0$. Odd spectral flow maps an open string with both endpoints on the
brane at $\psi = \psi_0$ to a string that begins at $\s = 0$ on
the $\psi_0$ brane but ends at $\s = \pi$ on the $-\psi_0$ brane, and {\em 
vice versa}.  The
two-brane system thus preserves the full spectral flow symmetry.  The
Hilbert space of the enlarged system has the structure
\beq
{\cal H} =   {\cal H}_{++} \oplus {\cal H}_{+-} \oplus {\cal H}_{-+}
\oplus {\cal H}_{--} \,,
\eeq
where, for example, ${\cal H}_{+-} $ is the Hilbert space of open
strings starting on the brane at $\psi = + \psi_0$ and ending on the
brane at $\psi = - \psi_0$, and similarly for the other summands.
Clearly ${\cal H}_{++} \cong {\cal H}_{--}$ and ${\cal H}_{+-} \cong
{\cal H}_{-+}$.

Each summand can be further decomposed as the sum of discrete and
continuous representations
of $\widehat{SL}(2,R)$.  As in the single-brane case, the symmetries
of the system give clues about how the representations in
various sectors of the theory
are related.  Spectral flow by an even amount is a
symmetry of each summand individually, while spectral flow by an odd
amount maps ${\cal H}_{++} \leftrightarrow {\cal H}_{+-}$ and ${\cal
H}_{--} \leftrightarrow {\cal H}_{-+}$.  For example, the action of
spectral flow with $w=1$ on the discrete representations of ${\cal
H}_{++}$ and ${\cal H}_{+-}$ is given
by
\begin{equation}
\label{diagram}
\begin{array}{ccccccccccc}
\cdots & \hat {\cal D}^{+,-2}_{++,j} &  & \hat {\cal D}^{+,-1}_{++,j} & &
\hat {\cal D}^{+,0}_{++,j} & &  \hat {\cal D}^{+, 1}_{++,j} & & \hat
{\cal D}^{+,2}_{++,j} & \cdots \\
 & & \nearrow \! \! \! \! \! \! \searrow & &
\nearrow \! \! \! \! \! \! \searrow & &
\nearrow \! \! \! \! \! \! \searrow & &
\nearrow \! \! \! \! \! \! \searrow & &  \\
\cdots & \hat {\cal D}^{+, -2}_{+-,j} &  & \hat {\cal D}^{+,-1}_{+-,j} & &
\hat {\cal D}^{+,0}_{+-,j} & &  \hat {\cal D}^{+, 1}_{+-,j} & & \hat
{\cal D}^{+,2}_{+-,j} & \cdots
\end{array} \,.
\end{equation}
Extending the reasoning of section \ref{quantumcurved}, we can deduce
that the $\psi_0$ dependence of the density of states in the
even (odd) winding discrete sectors of ${\cal H}_{++}$ must be the same as the
$\psi_0$ dependence of the density of states in the odd (even)
winding discrete sectors of ${\cal H}_{+-}$. A similar statement holds for
${\cal H}_{--}$ and ${\cal H}_{-+}$, and for the density of states of
continuous representations.  Within each summand, target space PT symmetry 
mixes discrete representations with odd and even $w$, but preserves
the parity of $w$ of the continuous representations.  We can hence
argue further that the $\psi_0$ dependence of the density of discrete
states is the same for all winding sectors and all summands.
By contrast, the $\psi_0$ dependence of the density of states in
the continuous representations is different in the odd winding sectors and
the even winding sectors in each summand, and the dependence in ${\cal
H}_{++}$ is different from the dependence in ${\cal H}_{+-}$.  One
advantage of enlarging our system to include a second brane is that
the entire structure of the continuous sector of the single-brane Hilbert space
is encoded in the representations $\hat {\cal C}^{\a,0}_{++,j}$, $\hat
{\cal C}^{\a,0}_{+-,j}$ and their properties under spectral flow and the
target space PT symmetry.

Just as in the $SU(2)$ WZW model, we may further generalize to a
system of $AdS_2$ branes located at $\psi = \psi_1$ and $\psi =
\psi_2$.  The discussion is completely parallel to what has already
been said.  We begin by constructing basic unwound classical solutions
stretching from one brane to the other, and employ spectral flow to
generate the space of all classical solutions. The basic unwound timelike 
solutions are again
\beq
t = \a \t \,,\qquad \th = \a \s + \th_0 \,,\qquad \r = \r_0 \,,
\eeq
where $0 \le \a < 1$, and $\th_0$ and $\r_0$ are chosen so as to
satisfy the appropriate Dirichlet conditions.  The basic spacelike
solution is given in matrix form as
 \beq
  g=\left( \begin{array}{cc}
             (1+a b) \exp(\a\t) & a \exp(\a\s) \\
              b \exp(-\a\s) &  \exp(-\a\t)
    \end{array}\right),
 \eeq
where the integration constants $a$ and $b$ are likewise fixed to satisfy
the Dirichlet boundary conditions.  

Spectral flow by an even amount $w$ is a symmetry of the system.
Short string solutions in odd winding sectors may be constructed by
means of the PT symmetry of the target space, as explained in section
\ref{classicalcurved}.   To construct
long string solutions belonging to odd winding sectors, we
introduce a third $AdS_2$ brane at $\psi = - \psi_2$, and follow the procedure
described in section \ref{classicalcurved} for finding solutions 
stretching between the $\psi_1$ and $-\psi_2$ branes.  Spectral flow
by an odd amount $w$ then gives open string solutions stretching
between the $\psi_1$ and $\psi_2$ branes.  

Again, these classical constructions provide us with a reasonable
conjecture for the quantum Hilbert space of the two-brane system: that
it contains discrete representations $\hat {\cal D}^{+,w}_j$ for all
integers $w$ and $j$ satisfying ${1 \over 2} < j < {k - 1\over 2}$, as
well as continuous representations $\hat {\cal C}^{\a,w}_j$ for all
integers $w$, all $\a$ in the range $0 \le \a < 1$, and $j$ of the
form $j = {1 \over 2} + is$ for all real $s$.  

\subsection{The NCOS Limit}
\label{NCOS}

\def\a{\alpha}
\def\b{\beta}
\def\g{\gamma}
\def\Img{({\rm Im} \, \gamma)^2}
\def\D{\Delta}
\def\l{\lambda}
\def\s{\sigma}
\def\t{\tau}
\def\vt{\vartheta}

\subsubsection*{The WZW Model Action}

In the limit $\psi_0 \rightarrow \infty$, the $AdS_2$ brane approaches
the boundary of $AdS_3$. In this limit, the factor of
$\cosh^2 \psi$ in the metric \eq{metric} tends to suppress open string
fluctuations along the $AdS_2$ brane. However, the induced
electric field on the brane also grows exponentially with $\psi_0$, 
balancing the effect of the metric. In fact, as $\psi_0
\rightarrow \infty$, the electric field approaches its critial
value, and the theory resembles the noncommutative open strings
studied in \cite{ncos,sst}. Let us now make this more precise.

We work in a target space of Euclidean signature. Our first task
is to find a system of coordinates that is well adapted to our
problem.  In the Euclidean $AdS_2$ coordinates $(\psi, \w, t)$, 
the $AdS_3$ metric is 
\beq 
ds^2 = d\psi^2+\cosh^2 \psi \, (\cosh^2 \w dt^2 + d\w^2) \,, 
\eeq 
and the invariant 2-form that describes the NS-NS background takes the
form 
\beq 
{\cal F} = \left( \psi -\psi_0 + {\sinh 2\psi \over 2} \right) \cosh
\w \, d\w \wedge dt \, . 
\eeq 
It proves convenient to replace the $AdS_2$ coordinates $(\w, t)$
by a complex coordinate $\g$, defined such that the $AdS_2$ brane
worldvolume is covered by the upper half plane ${\rm Im} \, \g \ge 0$.  
The transformations are 
\bea
\tanh \tau &=& {|\gamma|^2 - 1 \over |\gamma|^2 + 1} \, ,\\
\sinh \w &=& - {{\rm Re}\, \gamma \over {\rm Im} \,  \gamma} \, . 
\eea 
In these coordinates, the metric and 2-form become 
\beq
ds^2 = d \psi^2 + \frac{\cosh^2 \psi}{\Img} d \g d\bar \g \, ,
\eeq 
\beq {\cal F} = \frac{1}{2} \left( \psi - \psi_0 +
\frac{\sinh 2\psi}{2} \right) \frac{d \g \wedge d \bar \g}{\Img} \, . 
\eeq 

Given these expressions for the metric and the 2-form, the WZW model 
Lagrangian takes the form
\beq
{\cal L} = {k\over 2\pi} \left[ 2 \del\psi \bar\del\psi + {\cosh^2
\psi \over \Img} \left( \del\gamma \bar\del\bar\gamma+
\bar\del\gamma \del\bar\gamma \right) - \left(\psi - \psi_0 +
{\sinh 2\psi \over 2}\right) {1 \over \Img} (\del\gamma
\bar\del\bar\gamma - \bar\del\gamma \del\bar\gamma) \right] \, .
\eeq 
The worldsheet coordinates are $(z, \bar z)$, and range over the upper
half plane; the worldsheet metric is Euclidean; and the area element
is $d^2z \equiv dz \, d \bar z$.

To take the limit $\psi_0 \rightarrow \infty$, we first redefine
$\psi \rightarrow \psi + \psi_0$; the new coordinate $\psi$ 
measures deviations from the brane at $\psi_0$. The Lagrangian for 
large $\psi_0$ is then 
\beq 
{\cal L} = {k\over\pi} \, \del\psi \bar\del\psi + {k
\over 2\pi\Img} \left [ {e^{2(\psi+\psi_0)} \over 2}
(\bar\del\gamma \del\bar\gamma) + {1 \over 2} (\del\gamma
\bar\del\bar\gamma + \bar\del\gamma \del\bar\gamma) - \psi
(\del\gamma\bar\del\bar\gamma - \bar\del\gamma \del\bar\gamma)
\right] \, . 
\eeq 
It is useful to introduce Lagrange multipliers $\beta$ and
$\bar\beta$, giving
\beq 
{\cal L} = {k\over 2\pi}\left[2  \, \del\psi \bar\del\psi +
\beta \bar\del\gamma + \bar\beta \del\bar\gamma - {2 \Img \over
e^{2(\psi + \psi_0)} + 2} \beta \bar\beta + {1 \over \Img}
\left(\left({1 \over 2} - \psi \right) (\del\gamma
\bar\del\bar\gamma - \bar\del\gamma \del\bar\gamma)\right) \right]
\, . 
\eeq 
We now redefine $\psi \to \psi/2$ and take the limit $\psi_0
\rightarrow \infty$, with the result
\beq 
\label{lagrangian} 
S = {k\over2 \pi} \int d^2z 
\left({1\over 2} \del\psi \bar\del\psi + \beta \bar\del \gamma +
\bar\beta \del \bar\gamma + 2{\psi - 1 \over (\gamma -
\bar\gamma)^2} \left(\del\gamma \bar\del\bar\gamma -
\bar\del\gamma \del\bar\gamma \right) \right) \, . 
\eeq   
The action \eq{lagrangian} resembles the action of noncommutative open 
strings in flat space. The only differences are that the metric in the
$(\gamma,\bar{\gamma})$ plane is that of $AdS_2$, and that there
is a coupling to the extra scalar field $\psi$, which obeys the
Dirichlet condition $\psi=0$ on the boundary.

\subsubsection*{The Spectrum}

Our next goal is to evaluate the Euclidean one-loop 
thermal partition function exactly
and read off the physical string spectrum. As in Appendix \ref{partition}, 
the partition function may be written as a sum over a complete set of classical
solutions and fluctuations about these solutions.   Solving the WZW
model equations of motion produces classical solutions, given as
functions of $(\psi, \g, \bar \g)$; the fluctuations about these solutions
can then be evaluated by the method of iterative Gaussians
\cite{gawedzki}. From the resulting expression for
the partition function, we can determine which $\widehat{SL}(2,R)$
representations appear in the WZW model spectrum, as discussed in
Appendix \ref{partition}.  The result is somewhat unsettling: we find
that the Hilbert space contains only the discrete representation 
$\hat {\cal D}_j^{w=0}$, for ${1 \over 2} < j < {k-1 \over 2}$.  

This is clearly wrong: the existence of sectors of the Hilbert space
with $w \ne 0$ is
guaranteed by the spectral flow symmetry.  We understand their absence
from the partition function calculation to mean that our set of classical
solutions was incomplete.  In particular, there seem to be
solutions---short and long wound strings---that are not easily 
expressible in the 
$(\psi, \gamma, \bar{\gamma})$ coordinates. It
would be desirable to find all of the classical solutions 
and perform a complete
computation of the one-loop free energy. Nevertheless, our
computation captures part of the physical spectrum.  Most
important, our result implies that the spectrum contains
discrete representations with ${1 \over 2} <j< {k-1 \over 2}$, which is a
significant fact that cannot be seen at the semiclassical level.
These bounds on $j$ are the same as the bounds that come out of the
$\psi_0 = 0$ partition function calculation described in Appendix
\ref{partition}.   This strongly suggests that, for all $\psi_0$, the
spectrum contains discrete representations obeying ${1 \over 2} < j <
{k-1 \over 2}$.

The one-loop worldsheet is the semi-annulus in the upper half
$z$-plane defined by the
identification $z \sim z e^{2 \pi t}$, where $t$ is a worldsheet modulus.
At finite temperature $T$, the $AdS_3$ time coordinate is made
periodic with period $1/T$; in the coordinates we are
using, this is accomplished by identifying $\g \sim  \g e^{1/T}$.  In
order that the Lagrangian be single-valued, we must also identify $\b
\sim \b e^{-1/T}$.  These identifications mandate the relations 
 \beq
 \label{periods}
  \g(z e^{2\pi t})=e^{n/T}\g(z) \,, \qquad \b(z e^{2\pi
t})=e^{-n/T}\b(z) \,,
 \eeq
for some integer $n$. 

We begin by finding classical solutions for $\g$ and $\psi$.  The
equation of motion for $\b$ is $\delbar \g = 0$.  Classical solutions
are of the form 
\beq
\g_{cl}(z) = r e^{i \th} z^{{n \over 2 \pi t T}} \,,
\eeq
where $r e^{i \th}$ is a complex constant.  The range of the coordinate $\g$ is
$\Im \g \ge 0$.  This necessitates 
\beq
\label{tbound}
t \ge {|n| \over 2 \pi T} \,,
\eeq
and also determines $\th$ in a manner detailed below.

Let us now expand about the classical solutions, defining fields
$\G$ and $B$ by 
\beq
\g = \g_{cl} (1 + \G) \,, \qquad \b = {B \over \g_{cl}} \,.
\eeq
The partition function then involves functional integrals over $B$ and
$\G$.  Up to a constant, the $SL(2,R)$-invariant functional measure for these 
fields is
\beq
 {\cal D}\left({e^{\p/2} \g_{cl}\over \Im \g_{cl}}\G \right)
 {\cal D}\left({e^{-\p/2} \, \Im \g_{cl}\over \g_{cl}} B \right)
 {\cal D}\left({e^{\p/2} \bar\g_{cl}\over \Im \g_{cl}}\bar\G \right)
 {\cal D}\left({e^{-\p/2} \, \Im \g_{cl}\over \bar\g_{cl}} \bar B \right) \,.
\eeq
Classically, this is equivalent to ${\cal D} \G \, {\cal D} B \, {\cal D}
\bar \G \, {\cal D} \bar B$, but quantum mechanically, a chiral anomaly
is present, which effectively shifts the Lagrangian by
\beq
-{2 \over \pi} \left[ \del \log \left( {e^{\psi/2} \g_{cl} \over \Im
\g_{cl}} \right)  \delbar \log \left( {e^{\psi/2} \bar \g_{cl}
\over \Im \bar \g_{cl}} \right) \right] \,.
\eeq
When this term is expanded out and added to \eq{lagrangian}, the
result is 
 \beq
\label{newlag}
{\cal L}  = {k-2\over 2\pi}\left({1 \over 2} \del\psi \bar\del\psi +
   2 {\psi - 1 \over  (\g_{cl} - \bar\g_{cl})^2} \del\g_{cl}
   \bar\del\bar\g_{cl}\right)+{k\over 2\pi} \left(B \bar\del \G +
 \bar B \del \bar \G \right)+ F(\G, \del \G, \bar \G, \delbar \bar \G)\, ,
 \eeq
where $F$ contains terms from the Taylor expansion of the last term in
\eq{lagrangian}.  As these terms make no contribution to the partition
function, we drop them in what follows.  
The effect of the chiral anomaly is therefore to shift the prefactor $k$ of 
the terms in $\psi$ to $k-2$.

Next we separate $\psi$ into its classical and fluctuating parts,
defining a new field $\phi$ by $\psi = \psi_{cl} + \phi$, where
$\psi_{cl}$ satisfies the equation of motion
\beq
\del \delbar \psi_{cl} = {2 \del \g_{cl} \delbar \bar \g_{cl} \over
(\g_{cl} - \bar \g_{cl})^2} \,
\eeq
derived from \eq{newlag}.  One solution is
\beq
\psi_{cl} = 2 \log \Im \g_{cl} + f(z) + \bar f (\bar z) \,,
\eeq
where $f$ is an arbitrary holomorphic function.  We can use the
freedom of the choice of $f$ to ensure that $\psi_{cl}$ be
single-valued and satisfy its Dirichlet boundary condition.  Thus we may take 
\beq
\psi_{cl} = \log \left[ {\Im \g_{cl} \over |\g_{cl}|} \right]^2 + c \,,
\eeq
where the constant $c$ is adjusted so that $\psi_{cl} = 0$ when
$\Im z = 0$.  Substituting in the form of $\g_{cl}$ and letting $z = x
\in {\bf R}^+$, we must have $c = - 2 \log (\sin \th)$.  Letting $z =
-x$, we must have $c = -2 \log (\sin ({n \over 2 T t} + \th))$.
Consistency then requires $|\sin \th| = |\sin ({n \over 2 T t} +
\th)|$, which is satisfied (for $n \ne 0$) if
\beq
\th = {\pi \over 2} \left(m - {n \over 2 \pi T t} \right) \,,
\eeq
for some integer $m$.  
Since $\g_{cl}(z)$ is constrained to lie in the upper half plane as
$z$ ranges over the worldsheet semi-annulus in the upper half
$z$-plane, we must have $0 \le \arg(\g_{cl}(z)) = \th + {n \over 2 \pi T
t} \arg(z) \le \pi$, whenever $ 0 \le \arg(z) \le \pi$.  This is
possible only if $m = 1$.  Thus $\th = {\pi \over 2} (1 - {n \over 2
\pi T t})$.   In summary, our classical solutions have the form
 \bea
 \label{soln1}
   \g_{cl}=r e^{i{\pi\over 2}\left(1-{n \over 2\pi
   T t}\right)}z^{n \over 2\pi Tt} \,,\\
 \label{soln2}  
   \psi_{cl}=2 \log{\left|\mbox{Im} \, \g_{cl}\over \g_{cl}
   \cos({n\over 4 T t})\right|}\, ,
 \eea
where $t>{|n| \over 2 \pi T}$ and $r > 0$.  

Integration by parts brings the action into the form
 \beq
S = \int d^2z \left[{k-2\over 4\pi}\left(\del\phi \bar\del\phi +
   (\psi_{cl}-2)\del\bar\del\psi_{cl}\right)+{k\over 2\pi} \left(B
 \bar\del \G + \bar B \del \bar \G \right)\right] \,.
 \eeq
The term involving $\psi_{cl}$ can can be easily integrated over the
worldsheet semi-annulus, giving $(k-2)n^2/8\pi T^2 t$. 

Now we evaluate the Euclidean partition function.  The partition function breaks
up as a sum ${\cal Z} = \sum_n {\cal Z}_n$ over sectors indexed by the integer
$n$ that appears in the classical solutions \eq{soln1} and
\eq{soln2}. Each ${\cal Z}_n$ contains an integral $\int_{|n|/2
\pi T}^{\infty} {dt \over t}$ over the worldsheet
modulus $t$; the range of integration is set by \eq{tbound}.  There is
also an integral over the modulus $r$, but this contributes only a
numerical factor.  Finally, we must perform the functional integrals
over $\phi$, $B$, $\G$, the worldsheet ghosts, and the internal
conformal field theory.  Upon taking careful account of the boundary
conditions for $B$ and $\G$, we find that the functional integral of
the $(B, \G)$ system is exactly canceled by the functional determinant
coming from the worldsheet ghosts.  The remaining functional integral
is over the free bosonic field $\phi$, which obeys the Dirichlet
boundary condition $\phi = 0$.  This is a standard functional
determinant; the result is $1/|\eta(it)|$.   Assembling all the pieces
together, we have
\beq
\label{zn}
{\cal Z}_n \sim \int_{|n|/ 2\pi T}^\infty {dt \over t} {1 \over |\eta(it)|}
\exp \left(- {(k -2) n^2 \over 8 \pi T^2 t} \right) q^{h - {c_{int}
\over 24}} D(h)
\,,
\eeq
where $q = e^{-2 \pi t}$, $h$ indexes the weight in the internal
conformal field theory, $D(h)$ is the degeneracy at weight $h$, and
$c_{int}$ is the central charge of the internal conformal field theory.

As in \cite{part2} and Appendix \ref{partition}, it is sufficient for
our purposes to
look at the $n = 1$ sector.   The central charges $c_{SL(2,R)}$ of the
$SL(2,R)$ conformal field theory and $c_{int}$ of the internal
conformal field theory must sum to $26$; since $c_{SL(2,R)} = 3 + {6
\over k-2}$, we have $c_{int} = 23 - {6 \over k -2}$.  Substituting
this expression into \eq{zn}, and expanding the factors in $1/\eta(it) = e^{2 \pi
t /24} / \prod_{m=1}^{\infty} (1 - e^{-2 \pi t m}) $ as geometric
sums, we may rewrite the integrand as a sum of terms containing
exponentials of the form  
 \beq
  \exp\left(-2\pi t(h+N-1)-{ k-2 \over 8\pi T^2 t}-{\pi t \over 2 (k-2) }
  \right) \,,
 \eeq
where $N$ is an non-negative integer. The dominant contibution to
the partition function comes from the saddle point of the exponent,
 \beq
  t_s={k-2 \over 2\pi T \sqrt{1+4(k-2)(N+h-1)}} \,.
 \eeq
The lower bound of the $t$ integral forces
 \beq
  {1 \over 2\pi T}<t_s<\infty \,,
 \eeq
which translates into the bounds
 \beq
 \label{ineq}
  0<N+h-1+{1\over 4(k-2)}<{k\over 4}-{1\over 2} \,.
 \eeq
This is identical to the inequality (79) of \cite{part1}, with $w =
0$.   It was shown in \cite{part1} that \eq{ineq} is equivalent to
the bounds ${1 \over 2} < j < {k-1 \over 2}$ on the $SL(2,R)$ spin.  By a chain of
reasoning reviewed in Appendix \ref{partition} in the context of the
corresponding straight brane calculation, we may conclude that the
physical open string spectrum at $\psi_0 = \infty$ contains the
unwound discrete representation $\hat {\cal D}^{+,w=0}_j$, for all $j$ 
obeying ${1 \over 2} < j < {k-1 \over 2}$. 

As we remarked above, the appearance of the bounds ${1 \over 2} < j <
{ k-1 \over 2}$ at $\psi_0 = 0$ and at $\psi_0 = \infty$ gives us cause to
believe that the same bounds hold for all $\psi_0$.  Nonetheless, this 
calculation is manifestly incomplete, since it misses the winding
sectors.  Presumably this is
because our choice of 
coordinates is well suited to describing only a limited subset of the
classical solutions.  It would be interesting to find the remaining
solutions and complete the calculation of the partition sum.

\section{Summary and Discussion}

We have studied the spectrum of open strings ending on $AdS_2$
branes in $AdS_3$ in an NS-NS background. Perturbative open
string theory on an $AdS_2$ brane is described by the $SL(2,R)$ WZW
model, subject to the boundary conditions \eq{glue} and
\eq{dirichlet}, which state 
that the worldsheet ends on the $AdS_2$ brane and satsfies Neumann
boundary conditions in the directions parallel to the brane. The
condition \eq{glue}
also guarantees that the boundary
condition preserves one copy of the $\widehat{SL}(2,R)$ current algebra.

Our study of the open string spectrum has been modeled on the
treatment of closed strings in $AdS_3$ in \cite{part1}.  The
basic idea is to begin by studying classical solutions of the WZW
model, beginning with the simplest solutions and building up
more complicated ones by isometries and spectral flow, and, having
compiled a complete catalogue of the classical
solutions, to conjecture the form of the quantum Hilbert space.  It
was shown in \cite{part2} that this method leads to a correct proposal
for the closed string WZW model Hilbert space.

We have applied this approach to open strings ending on 
$AdS_2$ branes.  As a warm-up and a useful point of comparison, we
first looked at $S^2$ branes in the 
$SU(2)$ WZW model.  It was known from conformal field theoretic analysis
\cite{ars1} that the Hilbert space of open strings stretched between two 
$S^2$ branes at $\psi_0=\pi n_1/k$ and $\psi_0=\pi n_2/k$ is the sum of irreducible
highest-weight representations of $\widehat{SU}(2)$, whose spin $j$ is
bounded as in \eq{j2}. Our analysis of classical string worldsheets 
precisely reproduced this inequality. The only property of the quantum Hilbert
space that could not be seen from our classical analysis was the
quantization condition
\begin{eqnarray}   2j + n_1 + n_2 + k \equiv 0 \hbox{ mod } 2 \,.
\end{eqnarray}

We next considered strings ending on
$AdS_2$ branes in $AdS_3$. We began in section \ref{flatbrane} by
studying the straight brane located at $\psi_0=0$.  Analysis of classical
solutions led us to conjecture that the Hilbert space of
the WZW model in the open string sector is the holomorphic square root
of the closed string Hilbert space. In particular, the spectrum
contains both short strings and long strings, and is invariant under 
spectral flow.  We proved this conjecture in Appendix B
by exactly evaluating the one-loop open string free
energy, as in \cite{part2}.

The situation of the brane with $\psi_0 > 0$ is more interesting.
The boundary condition \eq{dirichlet} preserves only spectral flow
with even $w$, so it was reasonable to expect differences in the
$\psi_0$ dependence of the spectra in the even and odd winding
sectors.  We found both short and long classical string solutions in all
winding sectors, but also an important difference between the short
and long strings.  Short string solutions with different winding
numbers can be mapped to one another, even when the difference in
their winding numbers is odd, by combining spectral flow by an even
amount and a PT transformation of the target space.
Since both even spectral flow and the PT transformation
are symmetries of the full quantum theory, we argued that the $\psi_0$
dependence of the density of states of the short string solutions must
be the same in all winding sectors.

On the other hand, we showed that the $\psi_0$ dependence of the
density of states of long string solutions is different in the odd and
even winding sectors.  We saw this difference explicitly by computing
the energy of odd and even winding {\em short} strings as a function
of the string size.  In both cases, the energy rises to an asymptote,
signaling the presence of a continuum above the asymptotic value. The
rate of approach to the asymptote differs, though, in the odd and
even winding sectors.  Further evidence for the different behavior of
long strings with odd and even winding number came from an analysis of
the divergence structure of the Euclidean partition function.

In the limit $\psi_0 \rightarrow \infty$, the induced electric field
on the worldvolume of the $AdS_2$ brane reaches its critical
value, producing noncommutative open string theory on $AdS_2$.
In section \ref{NCOS}, we calculated the worldsheet action for open
strings in this limit, and obtained a result similar to that of \cite{gomis}
for noncommutative open strings in flat space \cite{ncos,sst}.
We carried out a partial computation of the one-loop free energy in
this limit, using the method of quadratures, as in \cite{gawedzki}.

Our work has focused exclusively on $AdS_2$ branes, which preserve
one copy of the $SL(2,R)$ current algebra. There are other branes
in $AdS_3$, some of which break the current algebra symmetry
entirely \cite{mms}. 
It would be interesting to analyze the open string theory of
these branes, as well.

\bigskip
\bigskip
\centerline{\bf Acknowledgments}

It is a pleasure to thank Jaume Gomis and Harlan Robins for
enlightening conversations.
P.L., H.O., and J.P. thank the Institute for
Theoretical Physics, Santa Barbara, for hospitality.

This research was supported in part by Department
of Energy grant DE-FG03-92ER40701,
National Science Foundation grant PHY99-07949,
and the Caltech Discovery Fund.

\appendix
\noindent

\section{Coordinate Systems for $AdS_3$}
\label{coordinates}

The space $AdS_3$ is defined as the
hyperboloid
\beq
\label{hyper}
(X^0)^2 - (X^1)^2 - (X^2)^2 + (X^3)^2 = R^2 \,,
\eeq
embedded in ${\bf R}^{2,2}$.  The metric
\beq
ds^2 = -(dX^0)^2 + (dX^1)^2 + (dX^2)^2 - (dX^3)^2
\eeq
on ${\bf R}^{2,2}$ induces a metric of constant negative curvature on
$AdS_3$. The quantity $R$ that appears in \eq{hyper} is the {\em
anti-de~Sitter radius}; for convenience, we set $R=1$.  In addition, to avoid
closed timelike curves, we work not with the hyperboloid
\eq{hyper} itself, but with its universal cover.

The two coordinate systems we use most extensively are {\em global
coordinates} and {\em $AdS_2$ coordinates}.  The global coordinates
$(\r, \th, \t)$ are defined by
\beq
X^0 + i X^3 = \cosh \r \, e^{i t} \,, \qquad X^1 + i X^2 = -\sinh \r \,
e^{-i \th} \,.
\eeq
The range of the radial coordinate $\r$ is $0 \le \r < \infty$; the
angular coordinate $\th$ ranges over $0 \le \th < 2 \pi$; and the
global time coordinate $t$ may be any real number.
The $AdS_3$ metric in global coordinates is
\beq
ds^2 = - \cosh^2 \!\r \, dt^2 + d\r^2 + \sinh^2 \!\r \, d \th^2 \,.
\eeq
The $AdS_2$ coordinates $(\psi, \w, t)$ are particularly well adapted
to the $AdS_2$ branes we consider in sections \ref{flatbrane} and
\ref{curvedbrane}.  They are defined by
\beq
X^1 = \cosh \psi \sinh \w \,, \qquad X^2 = \sinh \psi \,, \qquad X^0 +
i X^3 = \cosh \psi \cosh \w \, e^{it} \,.
\eeq
All three $AdS_2$ coordinates range over the entire real line. In this
parametrization, the fixed $\psi$ slices have the geometry of
$AdS_2$.  The $AdS_3$
metric in $AdS_2$ coordinates takes the form
\beq
ds^2 =  d\psi^2  + \cosh^2 \psi \, (-\cosh^2 \w \, dt^2 + d\w^2) \,;
\eeq
the quantity in parentheses is the metric of the $AdS_2$ subspace at
fixed $\psi$.   The transformation between global and $AdS_2$
coordinates is
\beq
\sinh \psi = \sin \th \sinh \r \,,\qquad \cosh \psi \sinh \w = - \cos
\th \sinh \r \,.
\eeq
The global time $t$ is the same in both coordinate systems.

The space $AdS_3$ is the group manifold of the group $SL(2,R)$.  A
point in $AdS_3$  is given by the $SL(2,R)$ matrix
\beq
g = \left( \begin{array}{cc} X^0 + X^1 & X^2 + X^3 \\ X^2 - X^3 & X^0
- X^1 \end{array} \right) \,.
\eeq
In the global coordinate system,
\beq
g =  \left( \begin{array}{cc} \cos t \cosh \r -
\cos \th \sinh \r & \sin t \cosh \r + \sin \th \sinh \r \\ -\sin t
\cosh \r + \sin \th \sinh \r & \cos t \cosh \r + \cos \th \sinh \r
 \end{array} \right) \,.
\eeq

\section{A Partition Function Calculation of the Open String Spectrum}
\label{partition}

In this appendix, we explicitly verify the proposal presented in
section \ref{quantumflat} for the open string spectrum.
First, in section \ref{z}, we compute the worldsheet one-loop partition
function ${\cal Z}$ for open string theory on Euclidean $AdS_3$ at finite
temperature $1/\b$.  The partition function is proportional to the
single-particle contribution to the spacetime free energy, ${\cal Z} =
- \b F$. The free energy, in turn, can be written as
\beq
F = {1 \over \b} \sum_{s \in {\cal H}} \log \left(1 - e^{-\b E_s} \right) \,,
\eeq
where the sum is over states $s$ in the physical Hilbert
space ${\cal H}$ of single-string states, and $E_s$ is the energy of
the state $s$.\footnote{We work at zero chemical potential.}
By writing ${\cal Z}$ in the right form, we can
thus read off the spectrum of open strings in (Lorentzian) $AdS_3$. We show in
section \ref{spectrum} that the spectrum breaks up into a sum over
discrete states and an integral over a continuum, with energies
agreeing with the expressions found in section \ref{flatbrane}.
Moreover, we compute the density of states of the continuum.

Our calculation is patterned on the one done in \cite{part2} for closed
strings in $AdS_3$.  Especially in section \ref{z}, we emphasize here
those features that are novel in the open string case; the reader
seeking greater detail is directed to \cite{part2} and the references therein.

One point is worth clarifying at the outset. Though the free energy
whose form we undertake to calculate receives contributions only from
states in the physical Hilbert space of the string, our calculation is
sufficient to confirm our proposal for the spectrum of the WZW model.
The physical Hilbert space is the tensor product of a Hilbert space of
$AdS_3$ excitations and a Hilbert space associated with the
``internal'' manifold ${\cal M}$.  For our purposes, we can take the
spectrum of the internal conformal field theory to be arbitrary.  One
of the physical state conditions is $L_0 + h = 1$, where $L_0$ is the
zeroth Virasoro generator of the $AdS_3$ conformal field theory, 
and $h$ is a conformal weight in the conformal field theory on ${\cal
M}$.  This condition can be seen as parametrizing the spectrum of
$L_0$ in the WZW model.  The remaining physical state conditions, $L_n
+ L_n^{{\cal M}} = 0$, with $n \ge 1$, relate the Virasoro generators
in the $AdS_3$ and internal conformal field theories.  They can be
solved within the tensor product of an irreducible representation of
$\widehat{SL}(2,R)$ with some subspace of the internal conformal field
theory state space.  Therefore, given the physical string spectrum in 
$AdS_3$, it is possible to deduce how the Hilbert space of the WZW model is
decomposed into a sum of irreducible representations of
$\widehat{SL}(2,R)$.  This is why the one-loop free energy computation
below, though it is carried out in the physical string Hilbert space,
is nonetheless relevant to the spectrum of the WZW model.

\subsection{The One-loop Partition Function}
\label{z}

Our first business is to write the WZW action for Euclidean $AdS_3$ at
finite temperature and the boundary conditions appropriate to a flat
$AdS_2$ brane.  We define the coordinates $(v,\bar v, \phi)$ on
Euclidean $AdS_3$ by
\bea
\label{H3coord}
v &=& \sinh\rho \, e^{i\th} \,,\\ \bar v &=& \sinh\rho \, e^{-i\th} \,,\\
\phi &=& t-\log\cosh\rho \,, \eea where $(\r,\th,t)$ are global
coordinates.  The metric in these coordinates is \cite{gawedzki}
\beq ds^2= k \left( d\phi^2+(dv+vd\phi)(d\bar v+\bar v d\phi)
\right) \,, \eeq where $k$ is the square of the anti-de~Sitter
radius, and is identified with the level of the WZW model.
Euclidean $AdS_3$ is the coset manifold $SL(2,C)/SU(2)$; in the
coordinates \eq{H3coord}, a general element is written as \beq g =
\left(  \begin{array}{cc} e^\phi (1 + |v|^2) & v \\ \bar v &
e^{-\phi} \end{array} \right) \,. \eeq

Thermal $AdS_3$ is given by the identification
\beq
t \sim t+ \b \,,
\eeq
where $\b$ is the inverse temperature.  In the coordinates
\eq{H3coord}, this translates to
\beq
\phi \sim \phi + \b \,.
\eeq

The WZW action in the coordinates \eq{H3coord} is\footnote{The action
given here differs from the WZW model Lagrangian given in section
\ref{NCOS} by boundary terms that can be ignored only if the
straight brane boundary condition \eq{BC} holds.  If the $AdS_2$ brane is
curved, these terms must be included.  The action then becomes
considerably more complicated, and loses some of the special
properties that make possible the partition function calculation described below.} 
\beq
\label{Zaction}
S={k\over\pi}\int d^2z \left(\del\phi \, \bar\del\phi+
(\del\bar v+\del\phi \,\bar v)(\bar\del v+\bar\del\phi \, v)\right) \,.
\eeq
Throughout this calculation, we take the worldsheet to be Euclidean.
In addition, we alternate between real $(\s,\t)$ and complex conjugate
$(z,\bar z)$ worldsheet coordinates.  The relation between the two
sets is
\beq
z=\s+i \tau \,, \qquad  \del\equiv{\del \over \del z} \,.
\eeq
and similarly for $\bar z$ and $\bar \del$.

In the coordinates (\ref{H3coord}), the boundary conditions suitable
for open strings ending on a straight $AdS_2$ brane are
\bea
\label{BC}
v_2 &=& 0 \,, \\
\label{BC2}
\del_\s v_1 &=& 0 \,,\\
\del_\s \phi &=& 0 \,, \eea where $v_1$ and $v_2$ are,
respectively, the real and imaginary parts of $v$.

The one-loop open string partition function is obtained by
considering worldsheets with the topology of a cylinder.  As
usual, $0 \le \s \le \pi$.  The worldsheet is made cylindrical by
imposing the periodicity \beq \t \sim \t + 2 \pi t \,. \eeq Here
(and henceforth) $t$ is simply a modulus, and is not to be
confused with the global time coordinate.  The spacetime
periodicity $\phi \sim \phi+\b$ of thermal $AdS_3$ implies
 \beq
  \phi(\s,\tau+2\pi t)=\phi(\s,\tau)+\b n \,,
 \eeq
for some integer $n$.  If we define
 \beq
  u_n={n\tau\over 4\pi t} \,
 \eeq
and
\beq
  \hat\phi=\phi-2\b u_n \,,
 \eeq
then $\hat\phi$ is periodic in $\tau$, {\em i.e.}, $\hat\phi(\t + 2
\pi t) = \hat \phi(\t)$.  The WZW action may be written in terms of
$\hat\phi$ as
 \beq
\label{WZWaction}
  S = \frac{k \b^2 n^2}{8 \pi t} + \frac{k}{\pi} \int d^2z
  \left(|\del \hat \phi|^2 + |(\del - i u_n + \del \hat \phi)
  \vbar |^2 \right) \,.
 \eeq

The partition function for Euclidean $AdS_3$ is
 \beq
\label{Zn}
  {\cal Z}_n(\b) \equiv \int{\cal D} \hat \phi \, {\cal D}v \, {\cal D}\vbar
 \, e^{-S} \,,
 \eeq
summed over $n$.   The rest of this subsection is devoted to
 evaluating the functional integrals in \eq{Zn}.

The second term in parentheses in \eq{WZWaction} couples $\phi$, $v$,
and $\bar v$.  The field $\hat \phi$ can be disentangled from $v$ and $\bar
v$ using a standard chiral gauge transformation and the chiral anomaly formulae
familiar from the closed string calculation. This procedure is valid
in the open string case as well, since the
string boundary conditions are left invariant by the chiral
transformation.  The partition function then factorizes into a
functional integral over $\hat \phi$ and a functional integral over
$v$ and $\bar v$, multiplied by the constant $e^{-k \b^2 n^2 / 8 \pi
t}$ coming from the constant term in the action.
The $\hat \phi$ functional integral is standard \cite{polchinski},
and, up to normalization, yields
\beq
{\cal Z}_{\hat \phi} = { \b (k-2)^{1/2} \over t^{1/2} | \eta(it)| } \,,
\eeq
where $\eta$ is the Dedekind eta function.   The
remaining functional integral may be written as
\beq
{\cal Z}_v = \int {\cal D}v \, {\cal D} \bar v e^{-S_v} \,,
\eeq
where
 \beq
 \label{Sv}
   S_v = \frac{k}{\pi}\int d^2z |(\del - i u_n) \bar
   v|^2.
 \eeq
Integrating $S_v$ by parts gives
 \beq
 \label{Sv2}
  S_v = -\frac{k}{\pi} \left(\int_\Sigma d^2z \, \vbar \, (\del+i u_n)
  \, (\delbar+i u_n) \, v + \int_{\del\Sigma} d\bar z \, v_1 \,
  \del_\s v_2 - u_n\int_{\del\Sigma} d\bar z \, v_1 v_1 \right) \,,
 \eeq
where $\Sigma$ denotes the worldsheet cylinder. Note that the 
two boundary terms are pure imaginary.

Let us work on the bulk term in \eq{Sv2}. We begin by expanding $v_1$ and $v_2$
in a complete basis of functions.  The boundary conditions \eq{BC} and
\eq{BC2} dictate the expansions
 \ber
  v_1(\s,\tau) &=& \sum_{M\ge 0,N\in {\bf Z}} a_{MN} {1 \over \pi
  \sqrt{2 t}}\cos \, M \s \, \psi_N(\tau/t) \,,\\
  v_2(\s,\tau) &=& \sum_{M> 0,N\in {\bf Z}} b_{MN} {1 \over \pi \sqrt{2
  t}}\sin \, M \s \, \psi_N(\tau/t) \,,
 \eer
where $a_{MN}$ and $b_{MN}$ are real-valued coefficients, and
$\psi_N$ is defined to be $\cos \, (N\tau/t)$ for $N\ge 0$ and
$\sin \, (N\tau/t)$ for $N<0$.  It follows that
 \beq
\label{vexpansion}
  v=v_1+i v_2 = \sum_{M, N \in {\bf Z}} v_{MN} {1 \over \pi
  \sqrt{2 t}}e^{iM \sigma}\psi_N(\tau /t) \,,
 \eeq
where
 \ber
  a_{MN} &=& ({v_{MN}+v_{-MN}})/2 \,, \\
  b_{MN} &=& ({v_{MN}-v_{-MN}})/2 \,.
 \eer

The advantage of this rewriting is that  $e^{iM\s}\psi_N(\tau/t)$ is an
eigenfunction of the operator $\D_v = -{k \over \pi} (\del + i u_n)
(\bar \del + i u_n)$ that appears in the bulk term of \eq{Sv2}, with eigenvalue
$\lambda_{MN} = -\left((M + \uhat)^2 + (N/t)^2\right)$.  If we
substitute the expansion \eq{vexpansion} of $v$ into \eq{Sv2}, then,
expressing $S_v$ in terms of the worldsheet coordinates $(\s,\t)$, we can
immediately integrate over $\t$, to obtain
 \ber
  S_v &=&  {1 \over \pi} \int_0^\pi d\s \sum_{M',M,N \in {\bf Z}}
  v_{M'N} \, v_{MN} \, \lambda_{MN}\left[ \cos M'\s \, \cos M\s + \sin
M'\s \, \sin M\s \right. \nonumber \\ && \left. + \, i \, \left(\cos M'\s
\, \sin M\s -
  \sin M'\s \, \cos M\s  \right) \right]+\mbox{boundary terms} \,,
 \eer
up to a normalization factor.
Since $S_v$ is positive-definite, $v_{MN}$ and $\lambda_{MN}$ are real
constants, and the
boundary terms are pure imaginary, the imaginary part of the bulk term
must cancel with the boundary terms.  We
are then left with
 \ber
  S_v &=& {1 \over 2} \sum_{M',M,N \in {\bf Z}}v_{M'N}\, v_{MN} \,
  \lambda_{MN} \, (\delta_{M',M}+\delta_{M',-M} + \delta_{M',M}-
  \delta_{M',-M}) \nonumber \\ &=& \sum_{M,N \in {\bf Z}}v_{MN} \,
  v_{M,N} \, \lambda_{M,N} \,.
 \eer
The functional integral is a product of Gaussians, and may be
evaluated by standard methods.  Up to a constant,
 \beq
  {\cal Z}_v = \prod_{M,N \in {\bf Z}} {1 \over \sqrt{(M + 2 u_n)^2 +
  (N/t)^2}} \,,
 \eeq
which may be zeta-function regularized
\cite{raysinger} to give
 \beq
  {\cal Z}_v^{-1} = \left|e^{-4 \pi u_n^2 t} \frac{\vt_1(-2it u_n,
  it)}{\eta(it)}\right| \,,
 \eeq
where $\vartheta_1$ is a Jacobi theta function.

We have now obtained expressions for all of the factors entering into
${\cal Z}_n(\b;t)$.  The overall normalization of ${\cal Z}_n$ is fixed in
the usual way, by examining the infrared limit.  Putting everything
together, we obtain
 \ber
  {\cal Z}_n(\b;t) &=& {1\over 4\pi\sqrt{2t}} \frac{\b (k-2)^{1/2}
  e^{-k \b^2 n^2/8 \pi t} e^{4\pi u_n^2 t}} {|\vt_1(-2it u_n ,
  it)|} \nonumber \\ &=& {1 \over 4\pi\sqrt{2t}} \frac{\b (k - 2)^
  {1/2}}{\sinh(\b n/2)}\frac{e^{-(k-2) \b^2 n^2/8 \pi t} \,\,
  e^{\pi t/4}}{\left[ \prod_{m=1}^\infty (1 - e^{-2 \pi t m})
  (1-e^{-2 \pi t m} e^{\b n}) (1 - e^{-2 \pi t m}e^{-\b n})
  \right]} \,. \nonumber \\
 \eer

The partition function we have calculated is that of a conformal field
theory with Euclidean $AdS_3$ as its target space, but our physical
open string theory contains more: we must incorporate the contributions
of the $(b,c)$ ghosts as well as those of the ``internal'' conformal
field theory.  In addition, we must integrate over the worldsheet
modulus $t$.  When this is done, the partition function becomes
\ber \label{fullZ}
  {\cal Z}(\b) &=& \frac{\b (k-2)^{1/2}}{4 \sqrt{2} \pi}
  \sum_{m=1}^\infty \int_{0}^\infty \frac{dt}{t^{3/2}} e^{2 \pi t
  (1-\frac{1}{4 (k-2)})} \sum_{h} D(h)e^{-2 \pi t h} \frac{e^{-(k-2) \b^2 m^2/8
  \pi t}}{\sinh(\b m/2)} \times \nonumber \\ && \prod_{n=1}^\infty
  \left| \frac{1 - e^{-2 \pi t n}}{(1 - e^{-2 \pi t n + \b m})(1 -
  e^{-2 \pi t n - \b m})}\right| \,.
 \eer
Here $h$ indexes the weight in the internal conformal field
theory, and $D(h)$ is the degeneracy at weight $h$.

\subsection{The Spectrum}
\label{spectrum}

Having calculated the partition function \eq{fullZ}, we now massage it
into a form from which we can read off the spectrum.
We noted at the beginning of this appendix that the partition function
${\cal Z}$ is proportional to the the free energy
\beq
F = {1 \over \b} \sum_{s \in {\cal H}} \log \left(1 - e^{-\b E_s}
\right) = \sum_{m=1}^\infty \sum_{s \in {\cal H}} {1
\over m \b} e^{-m \b E_s} \,.
\eeq
The partition function is
likewise a sum over $m$ of a function of $m \b$.  It suffices, then,
to compare the $m=1$ terms of the two expressions.   In this
subsection, we verify that $E_s$, extracted from the identification
 \ber
  \sum_{s \in {\cal H}} \frac{1}{\b}e^{-\b E_s}
  &=& \frac{\b (k-2)^{1/2}}{4\sqrt{2} \pi} \int_{0}^\infty
  \frac{dt}{t^{3/2}} e^{2 \pi t (1-\frac{1}{4 (k-2)})} \sum_{h}
  D(h)e^{-2 \pi t h} \frac{e^{-(k-2) \b^2
  /8 \pi t}}{\sinh(\b /2)} \times
  \nonumber \\ && \prod_{n=1}^\infty \left| \frac{1 - e^{-2 \pi t n}}
  {(1 - e^{-2 \pi t n + \b })(1-e^{-2 \pi t n - \b })}\right| \,,
 \eer
agrees with the string spectrum proposed in section \ref{flatbrane}.

To aid us in carrying out the $t$ integral, let us introduce a new
 variable $c$, defined by
 \ber e^{-(k-2)\b^2/8 \pi t} &=& - \frac{8 \pi i}{\b}
  \left({2 t\over k-2}\right)^{3/2} \int_{-\infty}^\infty \,\, dc \,\, c \,\,
  e^{-\frac{8 \pi t}{k-2}c^2 + 2 i \b c}.
 \eer
As explained in \cite{part2}, the right-hand side of
(\ref{fullZ}) can be expressed as a summation of terms of the form
 \ber \label{right}
  && {-4i\over \b (k-2)} \int_{-\infty}^\infty \,\, dc \,\, c
  \int_{\frac{\b}{2 \pi (w+1)}}^{\frac{\b}{2 \pi w}} \,\, dt
  \,\nonumber \\ && \:\:\:\:\:\:\:\: \times \exp\left[-\b\left(q+w+
  {1\over 2}\right)+2ic\b - 2\pi t\left(h+N_w+\frac{4 c ^2 + \frac{1}{4}}
  {k-2}-\frac{w (w+1)}{2} - 1\right) \right] \nonumber \\
  &&= \frac{-2i}{\pi\b} \int_{-\infty}^\infty \,\, dc
  \,\, c \,\, \frac{\exp\left[2 i c \b - \b (q + w + \frac{1}{2})\right]}{-2 \pi
  (h + N_w + \frac{4 c^2 + \frac{1}{4}}{k-2} - \frac{w (w+1)}{2} -
  1)} \nonumber \\ && \:\:\:\:\:\:\:\: \times \left\{-\exp\left[-\frac{\b}{w}
  \left(h+N_w-1+\frac{4 c^2+\frac{1}{4}}{k-2}-\frac{w
  (w+1)}{2}\right)\right]
  \right. \nonumber \\ && \:\:\:\:\:\:\:\: \left. +
  \exp\left[-\frac{\b}{w+1}\left(h+N_w-1+\frac{4 c^2+\frac{1}{4}}{k-2}-\frac{w
  (w+1)}{2}\right)\right] \right\} \,
 \eer
where $w$ ranges over non-negative integers.
We can complete the square of the exponent in the first term (the
fourth line) of (\ref{right}) by letting $c=s+{i\over 4}(k-2)w$.
Let us think of the $c$ integral as an integration over a contour (as
 it happens, the real line) in the complex plane.  We may then shift the
 contour of integration in the first term of \eq{right} to $
c=s+\frac{i}{4}(k-2)w$, and the contour of integration in the second
 term to $c=s+\frac{i}{4}(k-2)(w+1)$, where $s$ in both cases runs
 over the real line.  In doing so, the contour of integration crosses
 some poles in the integrand, and the integral picks up the residues
 of these poles.  The residues of the poles
 from the first
term are partially cancelled by the residues of the poles from the
 second.  The net result of the contour shift is to pick up only the poles in the range
 \beq
  {(k-2)\over 4} w < {\rm Im} \; c < {(k-2)\over 4}(w+1) \,.
 \eeq
Their residues are
 \beq \label{residue}
  {1\over \b} \,\, \exp \left[-\b q - \b
  \left({1\over 2}+w+\sqrt{\frac{1}{4}+(k-2) \left( N_w+h-1-{1\over
  2}w(w+1) \right) }\right)\right] \,.
 \eeq
The coefficient of $-\b$ in the exponent is supposed to be the energy
 of a typical state in the discrete spectrum.  Considerations similar to
 those given in \cite{part1} for closed strings show that
 \eq{openenergy} (with the minus sign chosen) indeed takes the form
 \eq{residue} after the physical state conditions are imposed.  Our
 partition function calculation thus reproduces the discrete spectrum
 of open strings in the physical Hilbert space.

We now turn our attention to the $s$ integration.  It is
convenient to rearrange the sum in (\ref{right}) by redefining $w
\rightarrow w-1$ in the second term and by deforming the contours
in both terms to $c=s+\frac{i}{4}(k-2)w$. The result is
 \ber \label{sintegral}
  && {1\over 2\pi i\b}\int_{-\infty}^\infty\,\, ds\,\, \left( {4s
  \over (k-2)w}+i \right) \nonumber \\ && \times \left\{ \frac{
  {\rm exp} \left[-\b q-\b\left({kw\over 4}+{1\over w}\left(\frac{4s^2+
  {1\over 4}}{k-2}+N_{w-1}+h-1\right)\right)
  \right]} {\frac{1}{4}+i s-\frac{kw}{8}+\frac{1}{2
  w}\left(N_{w-1}+h-1+\frac{4s^2+\frac{1}{4}}{k-2}\right)} \right.
  \nonumber \\ && \left. -\frac{{\rm exp} \left[-\b q - \b \left(
  {kw\over 4}+{1\over w} \left(\frac{4s^2+\frac{1}{4}}{k-2}+N_{w}+
  h-1\right)\right) \right]} {- \frac{1}{4}+i s-\frac{kw}{8}+{1 \over 2
  w}\left(N_{w}+h-1+\frac{4s^2+{1\over 4}}{k-2}\right)}
  \right\} \,.
 \eer
Let us consider the third line of (\ref{sintegral}).  It can be shown
\cite{part2}
that summing over
terms of this type yields
 \beq
  {1\over 2 \pi i\b}\int_{-\infty}^\infty\, ds\, \left(i+\frac{4s}{(k-2)w}\right)
  \left(2 \, \log \epsilon +  \left(i+{4s \over w(k-2)}\right)^{-1}
  {d\over ds} \log\Gamma\left(\frac{1}{2} - 2 i s - \tilde M\right) \right)
  e^{-\b f(s)} \,,
 \eeq
where
 \bea
  \tilde M = {1 \over w}\left({4s^2+{1\over4}\over k-2}+\tilde N + h - 1\right)
  - \frac{kw}{4} \,, \\
  f(s) = \frac{kw}{4} + {1 \over
  w}\left(\frac{4s^2+\frac{1}{4}}{k-2}+\tilde N+h-1 \right) \,,
 \eea
$\tilde N = q w + N_w$,  and $\epsilon$ is a cutoff introduced to regularize a divergence
that arises in the sum.  Similarly, summing over terms in the form of
the
second line of (\ref{sintegral}) gives
 \beq {1 \over 2 \pi i \b} \int_{-\infty}^\infty \,\, ds \,\,
  \left(i+\frac{4s}{(k-2)w}\right) \left( 2 {\rm log} \epsilon -
  \left(i+{4s\over w(k-2)}\right)^{-1} \frac{d}{ds} {\rm log} \Gamma
  \left( \frac{1}{2} + 2 i s + \tilde M \right) \right) e^{-\b f(s)}.
 \eeq
Combining these results and making the change of variables
$s \rightarrow {s\over 2}$, we
find that \eq{sintegral} can be written in the form
 \beq
  \frac{2}{\b} \int_{0}^\infty \,\, ds
  \,\, \rho(s) {\rm exp} \left[ -\b E(s) \right] \,,
 \eeq
where
 \beq \label{Lden}
  \rho(s)=\frac{1}{2\pi}2 \, {\rm log}\epsilon + \frac{1}{2 \pi
  i}\frac{1}{2} \frac{d}{ds} {\rm log} \left( \frac{\Gamma({1 \over
  2} - i s + \tilde m) \Gamma({1 \over 2} - i s - \tilde
  m)}{\Gamma({1 \over 2} + i s + \tilde m) \Gamma({1 \over 2} + i s
  - \tilde m)} \right) \,,
 \eeq
 \ber \label{Lengy}
  E(s) &=& \frac{kw}{4} + {1 \over
  w}\left(\frac{s^2+\frac{1}{4}}{k-2}+\tilde N+h-1 \right) \,,\\
  \tilde m &=& {1\over w} \left( \frac{s^2+{1\over 4}}{k-2}+\tilde N
  + h - 1 \right) - \frac{kw}{4} \,.
 \eer
Here $\r(s)$ and $E(s)$ are the density of states and energy of the
long strings.  The expression \eq{Lengy} is exactly what we would
find if we imposed the physical state conditions on the form
\eq{openenergy} given in section \ref{classicalflat} for the long
string energy.

Thus, by analyzing the partition function, we
have reproduced our conjecture for the spectrum of the straight brane. The
result is summarized by writing the free energy summand $f(\b)$ as
 \beq
  f(\b)={1\over\b}\sum D(h,{\tilde N},w)\left[\sum_q e^{-\b E(q)}+\int ds \,
  \r(s) e^{-\b E(s)}\right] \,,
 \eeq
where $E(q)$, $E(s)$, and $\r(s)$ are the discrete state energy, the
continuum state energy, and the continuum density of states.

\noindent

\bibliography{ads2}
\bibliographystyle{ssg}

\end{document}